\documentclass{article}

\usepackage{arxiv}

\usepackage[utf8]{inputenc} 
\usepackage[T1]{fontenc}    
\usepackage{hyperref}       
\usepackage{url}            
\usepackage{booktabs}       
\usepackage{amsfonts}       
\usepackage{nicefrac}       
\usepackage{microtype}      
\usepackage{lipsum}		
\usepackage{graphicx}
\usepackage{natbib}
\usepackage{doi}
\usepackage{amsmath}
\usepackage{algpseudocode}
\usepackage{algorithm}

\title{A Unified Framework for Multiple-Try Metropolis: Construction and Empirical Benchmarks}

\author{ \href{https://orcid.org/0000-0002-5691-0519}{\includegraphics[scale=0.06]{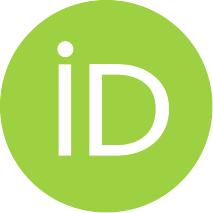}\hspace{1mm}Renny Doig} \\
	Department of Mathematics\\
	Simon Fraser University\\
	Burnaby, BC \\
	\texttt{rennyd@sfu.ca} \\
	\And
	\href{https://orcid.org/0000-0002-8509-7985}{\includegraphics[scale=0.06]{orcid.pdf}\hspace{1mm}Liangliang Wang} \\
	Department of Statistics and Actuarial Science\\
	Simon Fraser University\\
	Burnaby, BC \\
	\texttt{liangliang\_wang@sfu.ca} \\
}

\renewcommand{\shorttitle}{A Unified Framework for Multiple-Try Metropolis}

\hypersetup{
pdftitle={A template for the arxiv style},
pdfsubject={q-bio.NC, q-bio.QM},
pdfauthor={David S.~Hippocampus, Elias D.~Striatum},
pdfkeywords={First keyword, Second keyword, More},
}

\begin{document}
    \maketitle
    
    \begin{abstract}
    The multiple-try Metropolis (MTM) algorithm uses a compound proposal with multiple candidate draws to improve local sampling efficiency. While several methodological works have continued to develop MTM and the multi-candidate mechanism that characterizes it, the literature lacks a unified comparison of these components. This paper presents a structured formulation of MTM within the involutive MCMC framework, providing a principled approach for deriving valid acceptance probabilities based on the proposal mechanism. Through a comprehensive simulation experiment, we evaluate the impact of MTM configurations on non-Gaussian and multimodal target distributions. Our results reveal that while weight functions are a focus of several methodological developments, their impact on stationary sampling efficiency is secondary to the configuration of the proposal distribution. Furthermore, we find that while increasing the number of candidates enhances per-iteration efficiency, the realized performance gains are offset by computational overhead introduced by multiple candidacy unless parallelize computing is used. Our findings offer practical guidance for configuring an MTM algorithm for complex and non-Gaussian targets.
    \end{abstract}
    
    \keywords{Involutive Markov chain Monte Carlo \and Component-wise sampling \and Locally balanced proposals \and Bayesian computation}

\section{Introduction}\label{sec-intro}

Markov chain Monte Carlo (MCMC) is a broad class of algorithms used to sample from intractable probability distributions. Samples are generated by iteratively applying a Markovian transition kernel that is derived from a tractable proposal distribution. A fundamental MCMC algorithm is the Metropolis-Hastings (MH) algorithm \citep{hastings1970,metropolis1953}, which constructs a reversible transition kernel from a proposal distribution and an acceptance probability. There have been many subsequent developments in the MCMC literature that use this basic formula of pairing a proposal distribution and an acceptance probability. A unifying framework for many such methods has been presented in involutive MCMC, a general formula for MCMC based on a deterministic transition kernel \citep{neklyudov2020involutivemcmc}. This involutive MCMC can be used to frame MH as well as many of the algorithms that have descended from it.

One such descendant of MH is the multiple-try Metropolis (MTM) algorithm \citep{liu2000mtm}. MTM extends the basic formula of MH by substituting the single proposal with a compound proposal comprising multiple weighted candidates. Intuitively, generating a pool of candidates at each iteration increases the probability of identifying high-density regions, thereby improving mixing efficiency over standard MH. The original MTM algorithm has been occasionally, but consistently extended since it was initially proposed. These extensions can be roughly divided into those modifying the proposal distribution(s) \citep{casarin2013interactingMTM,yang2019componentwise} and those modifying the weight functions that are used for candidate selection \citep{gagnon2023localbalancing,pandolfi2014genMTMpap,pandolfi2010genMTMcon,yang2019componentwise}, although often both are modified in some way. Additionally, adaptation has been integrated into MTM frameworks to improve sampling efficiency \citep{fontaine2022adaptive,yang2019componentwise}.

In the academic literature on Bayesian computation, methodology papers typically use simulation to demonstrate specific use cases, often eschewing broader comparisons due to the prohibitive computational cost of large-scale Bayesian benchmarks. Comprehensive simulation studies are an important tool in developing a thorough understanding of methodology \citep{chipman2022sim,gelman2002graphs,harwell2017,skrondal2000}. While \citet{martino2018review} provided a valuable review connecting MTM to delayed rejection and particle MCMC, their study did not systematically evaluate MTM configurations. Consequently, a comprehensive experiment evaluating variations of the MTM algorithm remains absent from the literature.

In this paper, motivated by involutive MCMC, we present a constructive procedure for deriving acceptance probabilities in reversible Metropolis algorithms (Section \ref{sec:procedure P1}). We then apply this procedure to MTM and use this as the basis of a unified framework for MTM within which we consolidate several of its extensions (Section \ref{sec:mtm P1}). These extensions are evaluated systematically in a simulation experiment using a full factorial design (Section \ref{sec:simulation P1}). In this experiment we consider multiple types of target distributions, focusing on those exhibiting highly non-Gaussian features, multiple modes, as well as those arising from Bayesian data analysis problems.

\section{Constructing Reversible Kernels via Involutive MCMC}
\label{sec:procedure P1}

Suppose that we want to sample from a probability distribution $\pi$ that is defined for some state space $\mathcal X$. Reversible MCMC methods designed in the style of Metropolis and Hastings \citep{metropolis1953, hastings1970} simulate a Markov chain from a transition kernel composed of a proposal distribution $T(\mathbf y|\mathbf x)$ and an acceptance probability $a(\mathbf x,\mathbf y)$. Often the validity of these algorithms is established by showing that the detailed balance condition is met: $\pi(\mathbf x)T(\mathbf y|\mathbf x)a(\mathbf x,\mathbf y)=\pi(\mathbf y)T(\mathbf x|\mathbf y)a(\mathbf y,\mathbf x)$. 
Satisfying this condition ensures that the stationary distribution of the Markov chain is the target distribution, $\pi$.

Involutive MCMC \citep{neklyudov2020involutivemcmc} provides a general framework that allows many MCMC methods to be cast as a deterministic Metropolis algorithm, combining deterministic proposals \citep{tierney1998dettrans} with variable transformations \citep{geyer1994simulation,green1995jacobian}. In this section, we adapt the involutive MCMC framework to establish a constructive procedure for deriving valid acceptance probabilities from a given proposal mechanism. 
Because this approach guarantees reversibility with respect to $\pi$, it can be used to both unify existing and design novel MCMC algorithms. We apply this procedure to MTM, deriving its acceptance probability and placing existing MTM algorithms within a unifying framework.

\subsection{Connecting Deterministic Metropolis to Acceptance Probabilities}
\label{sec:derivation P1}

Let $\tilde\pi$ be a probability distribution over an extended measurable space $\tilde{\mathcal X}$. As will be illustrated later in this section, this space is ``extended'' in the sense that $\mathcal X\subset\tilde{\mathcal X}$. This state space may contain both continuous and discrete components, denoted $\mathcal X_C$ and $\mathcal X_D$, respectively. A {\it deterministic Metropolis transition kernel} \citep{tierney1998dettrans,neklyudov2020involutivemcmc} is constructed using a deterministic transformation $g: \tilde{\mathcal X} \to \tilde{\mathcal X}$ that is an involution, meaning $g(g(\tilde{\mathbf x})) = \tilde{\mathbf x}$. The transition kernel operates in two steps:

\begin{enumerate}
    \item A proposal is generated by applying the involution $g$ to the current state: $\tilde{\mathbf y}=g(\tilde{\mathbf x})$.
    \item The chain transitions to state $\tilde{\mathbf y}$ with probability 
    \begin{equation}\label{eqn:general_ratio_C_D}
a(\tilde{\mathbf x}) = 1\land \frac{\tilde\pi(g(\tilde{\mathbf x}))}{\tilde\pi(\tilde{\mathbf x})}|J(g_C(\tilde{\mathbf x}_C))|,\end{equation}
where $J(g_C(\tilde{\mathbf x}_C))$ is the Jacobian of the continuous component of the transformation. Otherwise, the chain remains in state $\tilde{\mathbf x}$ with probability $1-a(\tilde{\mathbf x})$.
\end{enumerate}

This deterministic kernel allows us to derive valid acceptance probabilities. Suppose an MCMC algorithm targets a distribution $\pi(\mathbf x)$ by generating a proposal $\mathbf y$. We define $\tilde\pi$ as the joint distribution over $\mathbf x$, $\mathbf y$, and any auxiliary variables generated during the proposal step. To ensure the resulting Markov chain targets the correct stationary distribution, the following condition must hold:

\paragraph*{Assumption 1} (Marginality). The target distribution $\pi(\mathbf x)$ must be recoverable as a marginal distribution of the extended joint distribution $\tilde\pi(\tilde{\mathbf x})$.

If Assumption 1 is satisfied, a sample from $\pi$ can be trivially extracted from a sample of $\tilde\pi$. By rewriting a probabilistic Metropolis method as a deterministic Metropolis kernel on the extended space, we can systematically derive its acceptance probability using the following five-step procedure:

\begin{enumerate}
    \item {\bf Define the extended state space}, $\tilde{\mathbf x}$, to include all random variables generated in a single iteration and specify the joint probability distribution over this space, $\tilde\pi(\tilde{\mathbf x})$.
    \item {\bf Define the involution}, $g(\tilde{\mathbf x})$, which represents the deterministic transformation applied to the extended state space if the proposal is accepted.
    \item {\bf Compute the Jacobian} of the continuous component of the involution, $J(g_C(\tilde{\mathbf x}_C))$.
    \item {\bf Verify that the target distribution}, $\pi(\mathbf x)$, is a marginal distribution of the extended joint distribution, $\tilde\pi(\tilde{\mathbf x})$.
    \item {\bf Derive the acceptance probability} by substituting these components into Equation \ref{eqn:general_ratio_C_D}.
\end{enumerate}

To illustrate this procedure, consider the standard Metropolis-Hastings (MH) algorithm targeting a stationary distribution $\pi$. Let the current state of the Markov chain be $\mathbf x$ and suppose a proposal $\mathbf y$ is generated from the distribution $T(\cdot|\mathbf x)$. The extended state space corresponding to this proposal is $\tilde{\mathbf x}=[\mathbf x, \mathbf y]$. This definition assigns the state of the Markov chain to the first coordinate of the extended space. The joint distribution over this extended space is $\tilde\pi(\tilde{\mathbf x})=\pi(\mathbf x)T(\mathbf y|\mathbf x)$. Accepting $\mathbf y$ as the next state would place $\mathbf y$ in the first coordinate, so the corresponding deterministic transformation is $g([\mathbf x,\mathbf y])=[\mathbf y,\mathbf x]$, which is trivially an involution. Because this transformation is a simple permutation, the Jacobian of its continuous component has an absolute value of 1. Furthermore, we can easily verify that $\pi(\mathbf x)$ is a marginal distribution of $\tilde\pi(\tilde{\mathbf x})$: since 
 $T(\mathbf y|\mathbf x)$ is a valid probability density, integrating $\tilde \pi$ over $\mathbf y$ yields $\pi(\mathbf x)$. 
Finally, the ratio $r(\tilde{\mathbf x})$ is computed as
\begin{equation*}
    r(\tilde{\mathbf x})=\frac{\tilde\pi(g(\tilde{\mathbf x}))}{\tilde\pi(\tilde{\mathbf x})}|J(g_C(\tilde{\mathbf x}))|=\frac{\pi(\mathbf y)T(\mathbf x|\mathbf y)}{\pi(\mathbf x)T(\mathbf y|\mathbf x)}.
\end{equation*}
Thus, the proposal $\mathbf y$ is accepted with probability $a(\tilde{\mathbf x})=a(\mathbf x,\mathbf y)=1\land r(\tilde{\mathbf x})$, the standard MH acceptance ratio.

\subsection{Deriving the MTM Acceptance Probability}
\label{sec:mtm accept prob P1}

In contrast with the single proposal from $T(\cdot|\mathbf x)$ that MH uses, MTM generates multiple candidates within each proposal \citep{liu2000mtm}. The proposal used by MTM is a compound mechanism made up of multiple proposal candidates, a candidate selection step, and finally the generation of auxiliary variables required for the evaluation of the acceptance probability. In this section we apply the constructive procedure described in Section \ref{sec:derivation P1} to derive the MTM acceptance probability.

To illustrate the MTM transition kernel, consider the chain at current state $\mathbf x\in\mathcal X$. Draw $M\ge1$ candidate states, $\mathbf y_m\sim T_m(\cdot|\mathbf x)$, for $m=1,\ldots,M$. A single candidate $\mathbf y=\mathbf y_J$ is then selected with probability $p(J|\mathbf y_{1:M},\mathbf x)$ for $J\in\{1,\ldots,M\}$. Conditional on this selection, $M-1$ auxiliary  ``reverse samples'' are generated:
\begin{equation*}
\mathbf x_m^*\sim
\begin{cases}
\delta_{\mathbf x}, & \text{if } m=J,\\
T_m(\cdot|\mathbf y), & \text{if } m\ne J,
\end{cases}
\end{equation*}
where $\delta_{\mathbf x}$ denotes the Dirac measure at $\mathbf x$.

The acceptance probability in Equation \ref{eqn:general_ratio_C_D} requires that we first define the extended state space, $\tilde{\mathbf x}=[\mathbf x,\mathbf y_{1:M},J,\mathbf x_{-J}^*]$, where $\mathbf x_{-J}^*=[\mathbf x_1^*,\ldots,\mathbf x_{J-1}^*,\mathbf x_{J+1}^*,\ldots,\mathbf x_M^*]$. If the chain is in its stationary distribution, the marginal distribution of the current state is $\pi(\mathbf x)$. The joint distribution over the extended space is the product of this marginal and the conditional distributions of the candidates, the selection index $J$, and the reverse samples:
\begin{equation*}
    \tilde\pi(\tilde{\mathbf x}) = \pi(\mathbf x)\left[\prod_{m=1}^M T_m(\mathbf y_m|\mathbf x)\right]p(J|\mathbf y_{1:M},\mathbf x)\prod_{m\ne J}T_m(\mathbf x_m^*|\mathbf y_J).
\end{equation*}

The involution operating on the extended space is $g(\tilde{\mathbf x})=[\mathbf y_J,\mathbf x_{1:J-1}^*,\mathbf x,\mathbf x_{J+1:M}^*,J,\mathbf y_{-J}]$, which swaps the current state $\mathbf x$ with the selected candidate $\mathbf y_J$, and the remaining candidates with the reverse samples, while leaving the index $J$ unchanged. Because this transformation is a permutation of variables, the Jacobian of its continuous component is 1 in absolute value. 

To verify that the target distribution $\pi(\mathbf x)$ is a marginal of $\tilde\pi(\tilde{\mathbf x})$, we integrate out the auxiliary variables. Integrating over the $M-1$ reverse samples $\mathbf x_{-J}^*$ yields:
\begin{align*}
    \int_{\mathcal X^{M-1}}\tilde\pi(\tilde{\mathbf x}) \mathrm d\mathbf x_{-J}^*
= & ~\pi(\mathbf x)\left[\prod_{m=1}^M T_m(\mathbf y_m|\mathbf x)\right]p(J|\mathbf y_{1:M},\mathbf x)\int_{\mathcal X^{M-1}}\prod_{m\ne J} T_m(\mathbf x_m^*|\mathbf y_J)\mathrm d\mathbf x_{-J}^* \\
= & ~ \pi(\mathbf x)\left[\prod_{m=1}^M T_m(\mathbf y_m|\mathbf x)\right]p(J|\mathbf y_{1:M},\mathbf x).
\end{align*}

Summing over the discrete selection index $J$ subsequently removes the selection probability $p(J|\mathbf y_{1:M},\mathbf x)$:
\begin{align*}
     \sum_{J=1}^M\left\{\pi(\mathbf x)\left[\prod_{m=1}^M T_m(\mathbf y_m|\mathbf x)\right]p(J|\mathbf y_{1:M},\mathbf x) \right\} 
    = & ~ \pi(\mathbf x)\left[\prod_{m=1}^M T_m(\mathbf y_m|\mathbf x)\right]\sum_{J=1}^Mp(J|\mathbf y_{1:M},\mathbf x) \\
    = & \pi(\mathbf x)\left[\prod_{m=1}^M T_m(\mathbf y_m|\mathbf x)\right]. 
\end{align*}
Finally, integrating over the joint candidate space $\mathbf y_{1:M}$ isolates the desired marginal distribution:
$\pi(\mathbf x)\int_{\mathcal X^M}\left[\prod_{m=1}^M T_m(\mathbf y_m|\mathbf x)\right]d\mathbf y_{1:M}  = \pi(\mathbf x).$

We derive the acceptance probability for the general MTM algorithm by evaluating Equation \ref{eqn:general_ratio_C_D}:
\begin{align*}
    \frac{\tilde\pi(g(\tilde{\mathbf x}))}{\tilde\pi(\tilde{\mathbf x})} &= \frac{\pi(\mathbf y_J)\left[\prod_{m=1}^MT_m(\mathbf x_m^*|\mathbf y_J)\right]p(J|\mathbf x_{1:M}^*,\mathbf y_J)\prod_{m\ne J}T_m(\mathbf y_m|\mathbf x)}{\pi(\mathbf x)\left[\prod_{m=1}^MT_m(\mathbf y_m|\mathbf x)\right]p(J|\mathbf y_{1:M},\mathbf x)\prod_{m\ne J}T_m(\mathbf x_m^*|\mathbf y_J)} \\
    &= \frac{\pi(\mathbf y_J)T_J(\mathbf x|\mathbf y_J)p(J|\mathbf x_{1:M}^*,\mathbf y_J)}{\pi(\mathbf x)T_J(\mathbf y_J|\mathbf x)p(J|\mathbf y_{1:M},\mathbf x)}.
\end{align*}
Thus, the selected proposal $\mathbf y_J$ is accepted with probability
\begin{equation}\label{eqn:general accept prob}
    a(\tilde{\mathbf x})=1\land\frac{\pi(\mathbf y_J)T_J(\mathbf x|\mathbf y_J)p(J|\mathbf x_{1:M}^*,\mathbf y_J)}{\pi(\mathbf x)T_J(\mathbf y_J|\mathbf x)p(J|\mathbf y_{1:M},\mathbf x)},
\end{equation}
which matches the so-called generalized MTM acceptance probability established in the literature \citep{pandolfi2014genMTMpap}.

\section{A Unified Construction of MTM Algorithms}
\label{sec:mtm P1}

Constructing an MTM algorithm requires specifying two components: the candidate proposal distributions, $T_m(\cdot|\mathbf x)$, and the candidate selection probabilities, $p(J|\mathbf y_{1:M},\mathbf x)$. How these are chosen determines the MTM variant. A complete outline of the MTM transition kernel is provided in Algorithm \ref{algo:basic MTM}.

\begin{algorithm}[]
\caption{Multiple-Try Metropolis Kernel}
\label{algo:basic MTM}
\begin{algorithmic}[1]
\State \textbf{Input:} Target distribution $\pi$; current state vector $\mathbf x$. 
\State \textbf{Settings:} Number of candidates $M$; proposal distributions $\{T_m\}$; selection probability function $p(J|\cdot)$.
\For{$m = 1, \dots, M$}
    \State Draw candidate $\mathbf y_m \sim T_m(\cdot|\mathbf x)$.
\EndFor
\State Sample an index $J \in \{1,\dots,M\}$ with probability $p(J|\mathbf y_{1:M},\mathbf x)$.
\State Set the selected candidate $\mathbf y = \mathbf y_J$.
\For{$m = 1, \dots, M$}
    \If{$m = J$}
        \State Set $\mathbf x_m^* = \mathbf x$.
    \Else
        \State Draw reverse sample $\mathbf x_m^* \sim T_m(\cdot|\mathbf y)$.
    \EndIf
\EndFor
\State Compute the acceptance probability $a(\tilde{\mathbf x})$ using Equation \ref{eqn:general accept prob}.
\State Draw $U \sim \text{Unif}(0,1)$.
\If{$U < a(\tilde{\mathbf x})$}
    \State \Return $\mathbf y$
\Else
    \State \Return $\mathbf x$
\EndIf
\end{algorithmic}
\end{algorithm}

\subsection{Configuring the Proposal Distribution}

The choice of proposal distribution dictates how the MTM algorithm explores the state space. The most common configuration is a 
random-walk proposal \citep{casarin2013interactingMTM,fontaine2022adaptive,gagnon2023localbalancing,liu2000mtm,pandolfi2014genMTMpap,yang2019componentwise}, typically of the form $\mathbf y_m\sim\mathcal N(\mathbf x,\Sigma_m)$. One must specify how $\Sigma_m$ depends on the candidate index $m$. If $\Sigma_m=\Sigma$ for all $m=1,\ldots, M$, the $M$ candidates $\mathbf y_{1:M}$ are independent draws from a common distribution $T(\cdot|\mathbf x)$, a configuration we refer to as a \textit{homogeneous} proposal \citep{fontaine2022adaptive,gagnon2023localbalancing,liu2000mtm}. 
Alternatively, the proposal distributions can be distinct, yielding a \textit{heterogeneous} configuration. For example, assigning a sequence of variances (i.e.\ $\sigma_1<\cdots<\sigma_M$) in a univariate setting would, if selected well, effect both local exploitation and broader exploration.

While the preceding description implicitly assumes the candidates, $\mathbf y_m$, are generated over the full state vector (\textit{full-block} updates), MTM can also be implemented by proposing updates to each dimension sequentially. In this \textit{component-wise} approach \citep{yang2019componentwise}, each component $i \in \{1, \dots, d\}$ is updated based on scalar proposals $y_{i,m}\sim T_{i,m}(\cdot|x_i)$. Algorithm \ref{algo:MTM CW} outlines this procedure. Although component-wise algorithms are performed in $\mathcal O(d)$ for each candidate, their theoretical computation time becomes comparable to full-block proposals if the target distribution permits $\mathcal O(1)$ component-wise likelihood evaluations.

\begin{algorithm}[]
\caption{Component-wise Multiple-Try Metropolis Kernel}
\label{algo:MTM CW}
\begin{algorithmic}[1]
\State \textbf{Input:} Target distribution $\pi$; current state vector $\mathbf x$. 
\State \textbf{Settings:} Number of candidates $M$; scalar proposal distributions $\{T_{i,m}\}$; selection probability $p(J|\cdot)$.
\For{$i = 1, \dots, d$}
    \For{$m = 1, \dots, M$}
        \State Set $\mathbf y_m = \mathbf x$.
        \State Draw the $i$-th element $y_{i,m} \sim T_{i,m}(\cdot|x_i)$.
    \EndFor
    
    \State Sample an index $J \in \{1,\dots,M\}$ with probability $p(J|\mathbf y_{1:M},\mathbf x)$.
    \State Set the selected candidate $\mathbf y = \mathbf y_J$.
    
    \For{$m = 1, \dots, M$}
        \State Set $\mathbf x_m^* = \mathbf x$.
        \If{$m \ne J$}
            \State Draw reverse sample $x_{i,m}^* \sim T_{i,m}(\cdot|y_i)$ for the $i$-th element.
        \EndIf
    \EndFor
    
    \State Compute the acceptance probability $a(\tilde{\mathbf x})$ using Equation \ref{eqn:general accept prob}.
    \State Draw $U \sim \text{Unif}(0,1)$.
    \If{$U < a(\tilde{\mathbf x})$}
        \State Update $\mathbf x = \mathbf y$.
    \EndIf
\EndFor
\State \Return $\mathbf x$
\end{algorithmic}
\end{algorithm}

Adaptation has also been used to improve the performance of MTM algorithms. In the context of MTM, adaptation typically means dynamically updating the proposal parameters based on the history of the chain or the local topography of the target. This alleviates the burden of time-consuming manual tuning while improving sampling efficiency. A standard strategy uses an iteratively updated empirical covariance matrix \citep{andrieu2008,fontaine2022adaptive}. After a burn-in period of $n_0$ iterations, a running mean vector $\pmb\mu$ and covariance matrix $\Sigma$ are stored. Note that $\pmb\mu$ is only used to compute $\Sigma$ and is not the mean of the random-walk proposal. To ensure diminishing adaptation and preserve the ergodicity of the Markov chain, a decaying learning rate scales the incremental update at each iteration, for example $\gamma(n)=n^{-0.6}$ \citep{fontaine2022adaptive}. The resulting proposal distribution is $\mathcal N(\mathbf x, \Sigma \times 2.38^2/d)$, where $2.38^2/d$ is the optimal covariance scaling factor for a $d$-dimensional target \citep{roberts1997optimalscaling}. The pseudocode is provided in Algorithm S1 of the Supplementary Material.

Adaptation mechanisms can also be tailored to exploit the multiple-try structure. Consider component-wise proposals defined by standard deviations, $\sigma_{i,m}$, that are equally spaced on the $\log_2$ scale. If candidates are disproportionately selected from proposals at one end of the scale, then some proposals are either too local or too global for the problem. To address this, \citet{yang2019componentwise} introduced an adaptation scheme that adjusts the variance bounds to achieve a well-balanced selection rate across all $M$ proposals. This approach monitors the empirical selection rate, $S_{i,m}$, for each coordinate $i$ and candidate index $m$. To ensure reliable estimates of these frequencies, the evaluation occurs every $\beta$ iterations. At these iterations, adaptation is carried out with a decaying probability $P=\max(0.99^{a-1},a^{-1/2})$, where $a=(n-\beta)/\beta$. Boundary variances are adjusted based on their empirical selection rates. If the largest variance is selected too frequently ($S_{i,M} > 2/M$), $\sigma_{i,M}$ is doubled; if it is selected infrequently ($S_{i,M} < 1/(2M)$) and $\sigma_{i,M}/2 > \sigma_{i,1}$, it is halved. Symmetric adjustments are applied to the lower bound, $\sigma_{i,1}$. Whenever a bound is modified, the intermediate variances $\sigma_{i,2:M-1}$ are re-interpolated to maintain equal spacing on the $\log_2$ scale. These are bounded within a range $[2^\epsilon, 2^L]$ for practical reasons. The pseudocode is given in Algorithm S2 in the Supplementary Material, with a thorough analysis available in \citet{yang2019componentwise}.

Although the focus of this paper is single-chain MTM methods, the algorithm naturally extends to multi-chain configurations. For instance, \citet{casarin2013interactingMTM} introduced a framework featuring $K$ simultaneous, interacting chains that target a common distribution. In their formulation, the $k$-th chain is updated by drawing $M$ candidates from a proposal distribution $T_m^{(k)}(\cdot|\tilde f^{(k)}(\mathbf x))$. The interaction mechanism is governed by the mapping $\tilde f^{(k)}(\mathbf z)=\left(\mathbf x^{(1:k-1)},\mathbf z,\mathbf x^{(k+1:K)}\right)^T$, which substitutes the proposed state $\mathbf z$ into the $k$-th position of the joint state vector across all chains. The form of $\tilde f$ allows practitioners to specify communication and interaction dynamics among the parallel chains. Formally unifying such multi-chain architectures within the involutive MCMC framework is left to future research.

\subsection{Weight Functions and Candidate Selection}

While the acceptance probability derived in Section \ref{sec:mtm accept prob P1} does not prescribe a specific form for the selection probability $p(J|\mathbf y_{1:M},\mathbf x)$, the MTM literature mostly uses a weight-based approach for sampling candidates. Specifically, a weight function $u_m(\mathbf y_m,\mathbf x)$ is defined that assigns weight to each candidate and reverse sample, and the selection probabilities  are the normalized weights:
\begin{equation}\label{eqn:selection prob P1}
p(J|\mathbf y_{1:M},\mathbf x)=\frac{u_J(\mathbf y_J,\mathbf x)}{\sum_{m=1}^M u_m(\mathbf y_m,\mathbf x)}.
\end{equation}

In general, $u_m(\mathbf y_m,\mathbf x)$ needs only be positive on the support of $\pi$ \citep{pandolfi2014genMTMpap}. However, many authors use a more structured form \citep{liu2000mtm}:
\begin{equation}\label{eqn:weight restricted P1}
u_m(\mathbf y_m,\mathbf x)=\pi(\mathbf y_m)T_m(\mathbf x|\mathbf y_m)\lambda(\mathbf y_m,\mathbf x),
\end{equation}
where the scaling function $\lambda(\mathbf y_m,\mathbf x)$ is positive on the support of $\pi$ and symmetric with respect to $(\mathbf y_m,\mathbf x)$. Substituting this restricted weight function into Equation \ref{eqn:general accept prob} simplifies the acceptance probability to:
\begin{equation}\label{eqn:restricted accept prob P1}
a(\tilde{\mathbf x}) = 1\land\frac{\sum_{m=1}^M u_m(\mathbf y_m,\mathbf x)}{\sum_{m=1}^M u_m(\mathbf x_m^*,\mathbf y)}.
\end{equation}

A trivial choice is the constant scaling function, $\lambda\equiv 1$, which yields $u_m(\mathbf y_m,\mathbf x)=\pi(\mathbf y_m)T_m(\mathbf x|\mathbf y_m)$, the joint distribution of $(\mathbf y_m,\mathbf x)$ under stationarity \citep{liu2000mtm}. Two examples of non-constant scaling functions are the so-called proportional and importance scalings:
\begin{equation}\label{eqn:proportional weight P1}
\lambda(\mathbf y_m,\mathbf x)=\left(\frac{T_m(\mathbf y_m|\mathbf x)+T_m(\mathbf x|\mathbf y_m)}{2}\right)^{-1} \quad \text{\citep{liu2000mtm}}
\end{equation}
and
\begin{equation}\label{eqn:importance weight P1}
\lambda(\mathbf y_m,\mathbf x)=\left[T_m(\mathbf x|\mathbf y_m)T_m(\mathbf y_m|\mathbf x)\right]^{-\rho}~~~\text{\citep{liu2000mtm}},
\end{equation}
where $\rho$ is a tuning parameter. Both functions satisfy the symmetry condition, regardless of the form of the proposal distribution. If the proposal distribution $T_m$ is symmetric, the weight function corresponding to Equation \ref{eqn:proportional weight P1} simplifies to $u_m(\mathbf y_m,\mathbf x)=\pi(\mathbf y_m)$. In this case, candidate $m$ is selected with probability proportional to its target density, hence the term ``proportional'' weight. Similarly, if $\rho=1$ in Equation \ref{eqn:importance weight P1} and the proposal is symmetric, the weight simplifies to $\pi(\mathbf y_m)/T_m(\mathbf y_m|\mathbf x)$, drawing an analogy to importance sampling.

A fourth variation designed to encourage exploration is the jump distance scaling function:
\begin{equation}\label{eqn:jump distance weight P1}
\lambda(\mathbf y_m,\mathbf x)=T_m(\mathbf y_m|\mathbf x)^{-1}\|(\mathbf y_m-\mathbf x)\|^\alpha~~~\text{\citep{yang2019componentwise}},
\end{equation}
where $\alpha$ is a tuning parameter and $T_m$ is a symmetric proposal distribution. By explicitly incorporating the Euclidean distance between the candidate and the current state, greater weight is assigned to candidates making larger jumps. Empirical evidence suggests an optimal range of $\alpha\in(2,4)$ \citep{yang2019componentwise}.

More recently, a theoretical treatment of MTM has produced the so-called locally balanced weight function,
\begin{equation*}
u_m(\mathbf y_m,\mathbf x)=\sqrt{\pi(\mathbf y_m)},
\end{equation*}
which has been shown to accelerate convergence \citep{gagnon2023localbalancing}. Furthermore, if the proposal distributions are homogeneous, the resulting MTM kernel that approximates Hamiltonian Monte Carlo as $M\to\infty$. These five weight functions are summarized in Table \ref{tab:weight functions}.
\begin{table}[h!]
    \centering
    \begin{tabular}{|c|c|c|}
    \hline
        Weight function & Functional form & Source  \\
        \hline
        Constant & $\pi(\mathbf y_m)T(\mathbf x|\mathbf y_m)$ & \cite{liu2000mtm} \\
        Importance & $\pi(\mathbf y_m)/T(\mathbf y_m|\mathbf x)$ & \cite{liu2000mtm} \\
        Proportional & $\pi(\mathbf y_m)$ & \cite{liu2000mtm} \\
        Locally balanced & $\sqrt{\pi(\mathbf y_m)}$ & \cite{gagnon2023localbalancing} \\
        Jump distance & $\pi(\mathbf y_m)\|\mathbf y_m-\mathbf x\|^\alpha$ & \cite{yang2019componentwise} \\
        \hline
    \end{tabular}
    \caption{Weight functions discussed in Section \ref{sec:mtm P1}, specified in the form $u_m(\mathbf y_m,\mathbf x)$.}
    \label{tab:weight functions}
\end{table}

The general form for the weight can be exploited to construct an approximate MCMC method that trade exactness for computational efficiency. Evaluating the exact acceptance probability in Equation \ref{eqn:restricted accept prob P1} requires computing the target density $2M$ times per iteration. To reduce this computational bottleneck, \citet{pandolfi2014genMTMpap} proposed using a second-order Taylor expansion of the target log-density around the current state $\mathbf x$:
\begin{align*}
    \pi^*(\mathbf y_m,\mathbf x) &= \pi(\mathbf x)B(\mathbf y_m,\mathbf x), \\
    B(\mathbf y_m,\mathbf x) &= \exp\left[s(\mathbf x)^T(\mathbf y_m-\mathbf x)+\frac{1}{2}(\mathbf y_m-\mathbf x)^TD(\mathbf x)(\mathbf y_m-\mathbf x)\right], \\
    u_m^*(\mathbf y_m,\mathbf x) &= \pi^*(\mathbf y_m)T_m(\mathbf y_m|\mathbf x)\lambda(\mathbf y_m,\mathbf x),
\end{align*}
where $s(\mathbf x)$ and $D(\mathbf x)$ denote the gradient and Hessian of $\log\pi(\mathbf x)$, respectively. By substituting this approximation into the weight function, the true target distribution $\pi$ only needs to be evaluated at the current state $\mathbf x$ and the selected proposal $\mathbf y_J$. However, because $\pi^*$ is used to construct the kernel, the stationary distribution of the resulting Markov chain is $\pi^*$ rather than the target $\pi$.

\section{Simulation Experiments}
\label{sec:simulation P1}
To benchmark MTM performance, we conduct a full-factorial simulation study across three classes of target distributions: non-Gaussian targets, multimodal landscapes, and posterior distributions arising from parameter inference problems. While a previous review has explored select MTM configurations in the context of signal processing \citep{martino2018review}, our study provides a systematic evaluation of MTM settings, including recent developments since the last review, to fully map the interaction effects between algorithm design choices.

\paragraph*{Algorithm Design and Experiment Settings} All experiments vary three factors relating to the MTM algorithm design: the weight function, the form of the proposal distribution, and the number of candidate draws. We evaluate four weight functions (Table \ref{tab:weight functions}): proportional, importance, jump distance, and locally balanced. The proposal distributions are defined by their composition (homogeneous vs. heterogeneous) and the update scheme (full-block in Algorithm \ref{algo:basic MTM} vs. component-wise [CW] in in Algorithm \ref{algo:MTM CW}). We collectively refer to the combination of composition and update scheme as the ``proposal configuration''. This yields four distinct proposal configurations: Homogeneous-Full, Heterogeneous-Full, Homogeneous-CW, and Heterogeneous-CW. In all four settings, a Gaussian random-walk proposal is used with an adaptive (co)variance. Both Homogeneous-Full and Heterogeneous-Full use the standard Adaptive Metropolis procedure (Algorithm S1 of the Supplementary Material) \citep{fontaine2022adaptive,haario1999}, Homogeneous-CW employs a univariate version of the Adaptive Metropolis procedure, and Heterogeneous-CW uses the Balanced Selection Rate adaptation \citep{yang2019componentwise}.

The number of candidates ($M$) varies by experiment. For all configurations except Heterogeneous-CW (which requires $M \ge 2$ for its adaptation scheme), we include $M=1$ as a baseline to represent the Metropolis-Hastings algorithm. Note that multi-chain variations of the algorithm \citep{casarin2013interactingMTM} are excluded from this benchmark.

\paragraph*{Computing Resources}
For each experimental setting, 50 independent Markov chains are generated. To ensure a fair comparison of sampling efficiency, the algorithms are run for a fixed computational time budget rather than a fixed number of iterations. All simulations are executed on an Intel Gold 6148 Skylake @ 2.4 GHz CPU. The algorithms are implemented in Julia \citep{julia}, with post-processing and visualizations rendered in R \citep{r, ggplot2}. The complete source code to reproduce this study is available at \href{https://github.com/SFU-Stat-ML/multiple-try-metropolis-experiments}{https://github.com/SFU-Stat-ML/multiple-try-metropolis-experiments}.

\paragraph*{Automated Burn-in Detection}
Given the volume of chains generated across the factorial design, manual assessment of convergence is infeasible. Additionally, discarding a fixed proportion of iterations is inappropriate due to the highly variable mixing rates across different MTM configurations. To ensure that stationary performance is evaluated accurately, we apply an automated, ad-hoc burn-in procedure designed to identify a burn-in period that produces a stable mean estimate. All experiments use this procedure to process the resulting chains prior to visualisations, unless otherwise specified. Algorithm details can be found in Algorithm S3 in the Supplementary Material.

\subsection{Non-Gaussian Target Distributions}

In this simulation study, we define ``non-Gaussian'' targets as distributions exhibiting features, such as severe, state-dependent curvature, that cannot be well-approximated by a Gaussian distribution except at very local scales. Because of this, MCMC methods based on local moves, especially those relying on a Gaussian random-walk as a proposal, are forced to take very small steps to maintain a reasonable acceptance rate. Many modern MCMC methods struggle to overcome this difficulty when such features are present in the extreme. The following simulations investigate which MTM settings are best-suited for sampling from these non-Gaussian distributions. 

\begin{figure}[] 
\centering
\includegraphics[width=0.8\textwidth]{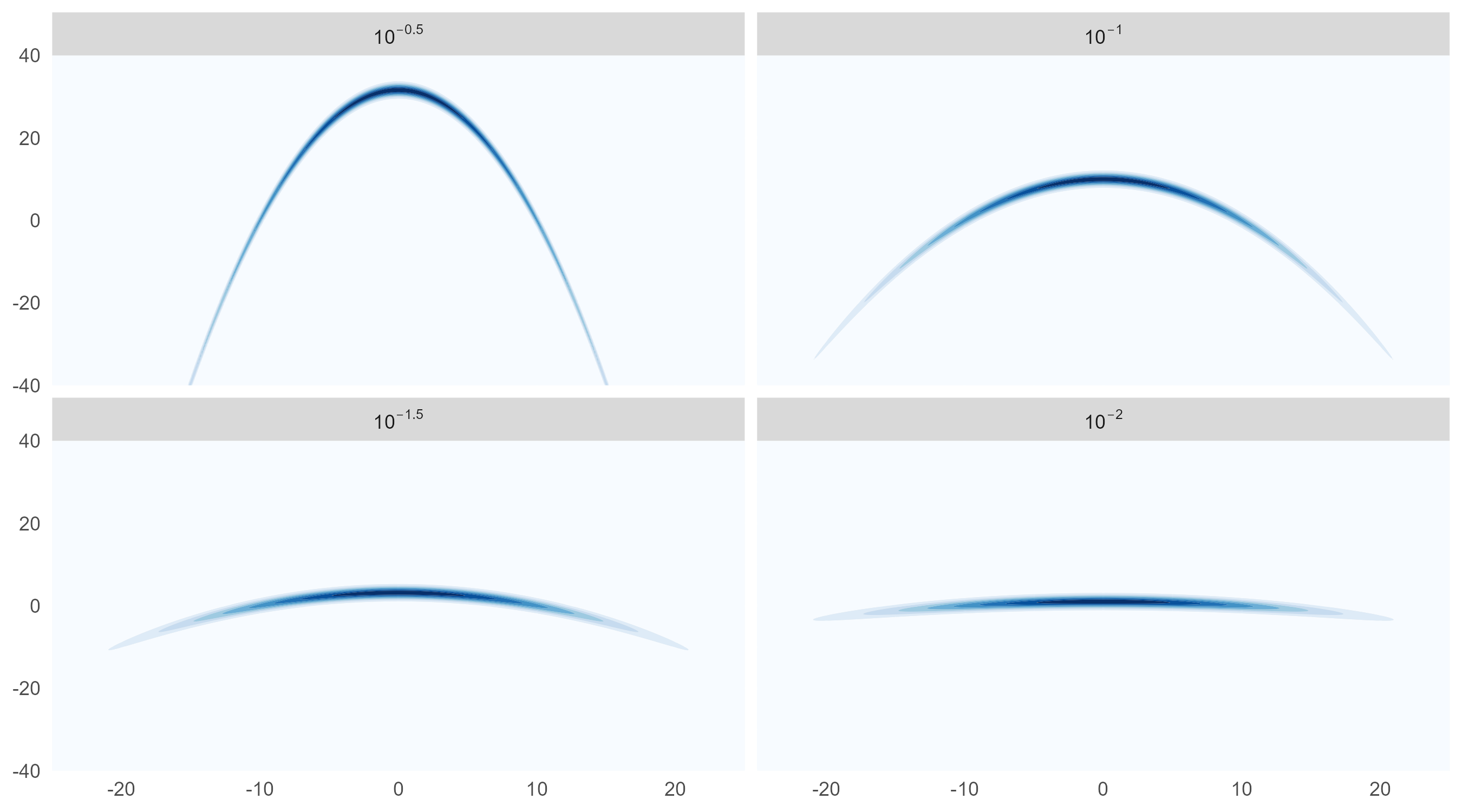}
\caption{Contours of the banana distribution's first two marginals. Panels correspond to the non-Gaussianity parameter $B$.}
\label{fig:banana example P1}
\end{figure}

\paragraph*{Banana Distribution} The first target is the ``banana distribution,'' a standard benchmark for evaluating adaptive MCMC algorithms due to its nonlinear, crescent-shaped contours \citep{andrieu2008,bironlattes2024automala,fontaine2022adaptive,haario1999,yang2019componentwise}. The distribution is defined by a transformation of a $d$-dimensional standard Gaussian vector $\mathbf x$ that results in the following unnormalized density:
$$\pi(\mathbf x) \propto \exp\left[-\frac{x_1^2}{200}-\frac{1}{2}\left(x_2+Bx_1^2-100B\right)^2-\frac{1}{2}\sum_{i=3}^d x_i^2\right].$$
The parameter $B$ controls the degree of non-Gaussianity, such that sampling difficulty increasing as $B$ grows. The resulting distribution is unimodal with contours taking a crescent shape when jointly plotting $(x_1,x_2)$, making the first two coordinates the hardest to sample. Alternative parameterizations can result in multiple coordinate pairs having non-Gaussian curvature \citep{fontaine2022adaptive}. We evaluate performance across a grid of non-Gaussianity, setting $\log_{10}B \in \{-2, -1.75, \ldots, -0.5\}$. Contours of the target for four values of $B$ are shown in Figure \ref{fig:banana example P1}.

We benchmark the four proposal configurations and four weight functions across candidate pool sizes $M \in \{1, 5, 10, 15, 20\}$, yielding 560 experiment settings. To measure sampling efficiency, we compute the multivariate effective sample size (mESS) \citep{vats2019mess} after running each algorithm for a fixed wall-clock budget of ten seconds.

Regarding the relative speeds of the algorithm settings, the full-block proposals were 
substantially faster per iteration than their component-wise counterparts. This is expected, as our implementation of the component-wise algorithm requires $d$ times the number of target density evaluations for each candidate, relative to the full-block algorithm. The runtime difference between Homogeneous-CW and Heterogeneous-CW was negligible, whereas the Heterogeneous-Full configuration was notably slower than Homogeneous-Full.

\begin{figure}[] 
\centering
\includegraphics[width=\textwidth]{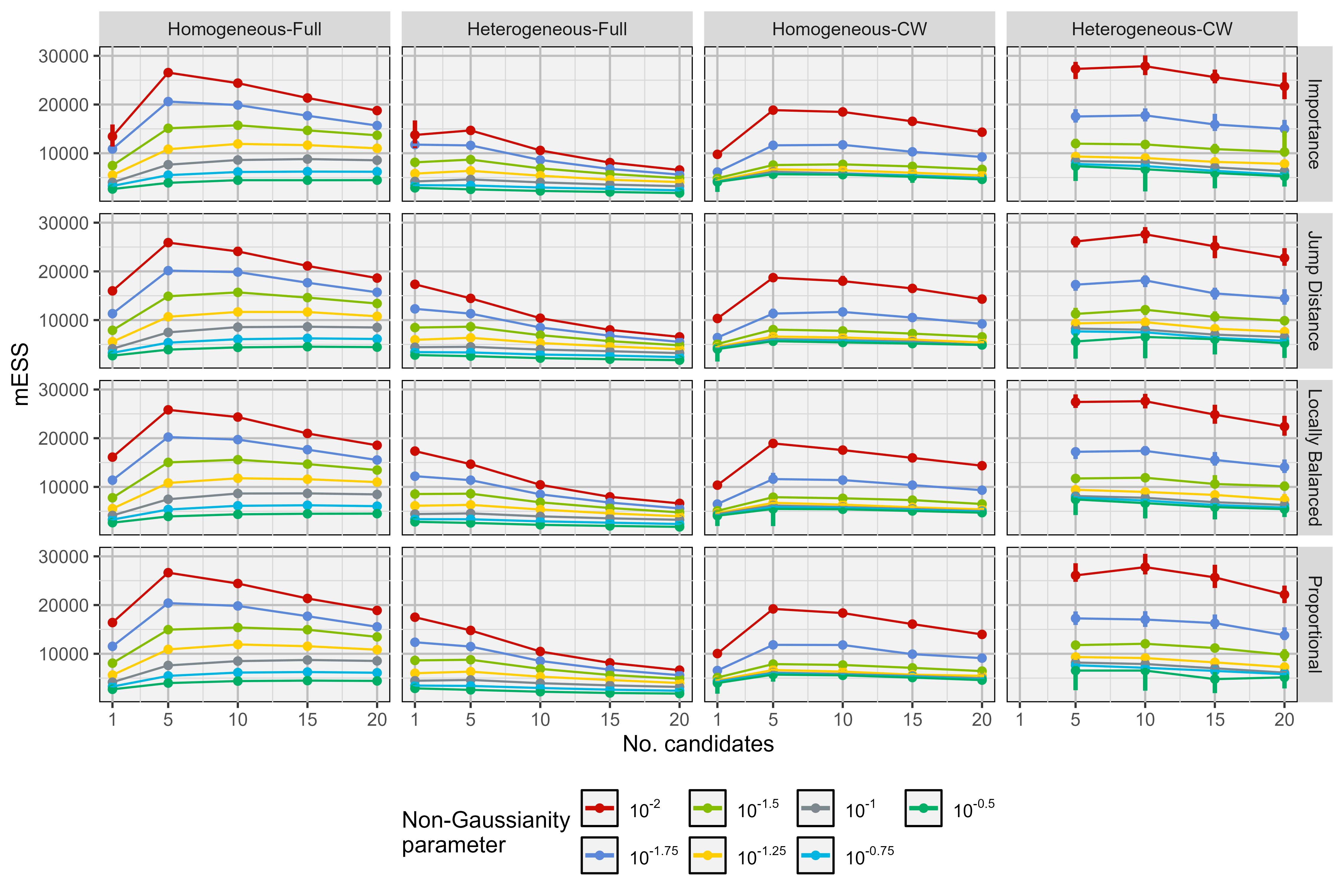}
\caption{Raw mESS for the banana distribution. The median (point) and 50\% interval (vertical line) over 50 independent runs. Column panels correspond to the proposal distribution configuration and row panels to the weight function.}
\label{fig:banana_raw}
\end{figure}

The distribution of raw mESS across the 50 independent runs is shown in Figure \ref{fig:banana_raw}. When comparing the MH baseline ($M=1$) to the MTM extensions ($M>1$), sampling efficiency using mESS improves in all configurations except Heterogeneous-Full.  However, this benefit diminishes at higher candidate counts: both Homogeneous proposals suffer performance degradation for $M>5$, while the Heterogeneous-CW proposal remains stable until $M>10$. The choice of weight function exhibits little observable impact on sampling efficiency.

\begin{figure}[] 
\centering
\includegraphics[width=\textwidth]{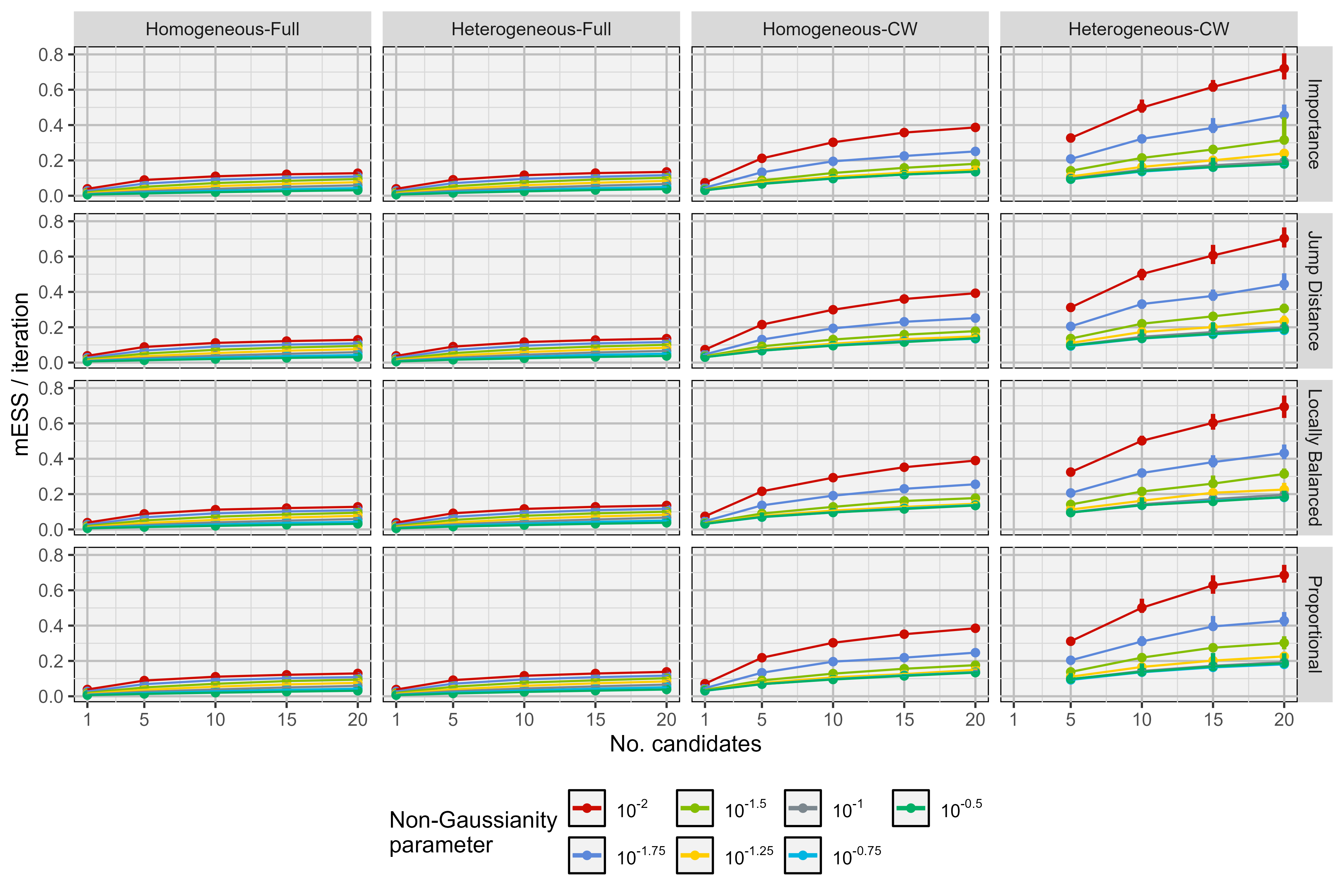}
\caption{mESS per iteration for the banana distribution. The median (points) and 50\% interval (vertical lines) over 50 independent runs are shown for each setting. Column panels correspond to the proposal distribution configuration and row panels to the weight function.}
\label{fig:banana_std}
\end{figure}

To isolate the sampling efficiency from computational overhead, Figure \ref{fig:banana_std} plots the mESS per iteration. This metric illustrates algorithm performance in an ideal scenario where the $M$ candidate evaluations are perfectly parallelized. Unconstrained by serial computational costs, all combinations of weight functions and proposal configurations demonstrate strictly increasing per-iteration mESS as $M$ grows. The component-wise proposals yield greater per-iteration efficiency than the full-block proposals, with Heterogeneous-CW outperforming Homogeneous-CW. As with the raw mESS results, the choice of weight function shows no significant effect on per-iteration gains.

\paragraph*{Neal's funnel} Neal's funnel is a $(d+1)$-dimensional target where the variance of the latter $d$ dimensions scales exponentially with the first dimension ($y$), creating a narrow, funnel-shaped geometry that impedes random-walk exploration \citep{neal2003slice,bironlattes2024automala, modi2024DRHMC}. The distribution specification is:
\begin{align*}
    y &\sim \mathcal N\left(0,3^2\right), & x_i &\sim \mathcal N\left(0,\left[e^{y/\beta}\right]^2\right) \quad \text{for } i=1,\ldots,d.
\end{align*}
Here, $\beta$ acts as an inverse scale parameter, controlling the steepness of the funnel. We evaluate the performance for $\log_{10}\beta^{-1} \in \{-0.5, -0.25, 0, 0.25, 0.5\}$. Algorithms are run for ten seconds across all 400 settings ($M \in \{1,5,10,15,20\}$). Because the exact mean of $y$ is known to be zero, performance is evaluated by the accuracy and Monte Carlo standard error of the empirical mean of the first marginal, $y$.

\begin{figure}[] 
\centering
\includegraphics[width=\textwidth]{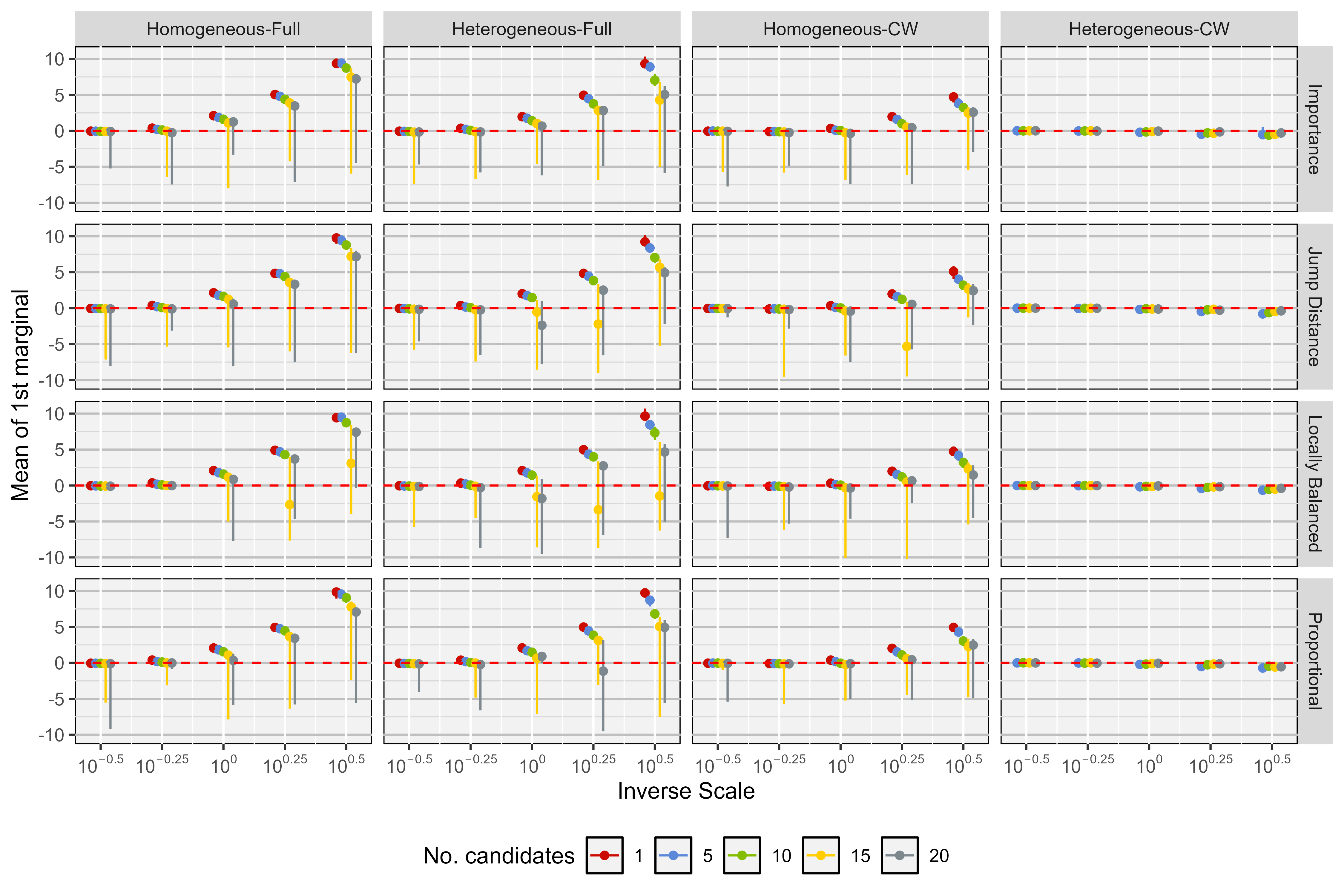}
\caption{Empirical mean of the first marginal ($y$) of Neal's funnel. Points represent the median and vertical lines the 50\% interval across 50 independent runs. The true mean is zero. Column panels correspond to the proposal configuration and row panels to the weight function.}
\label{fig:funnel_mean}
\end{figure}

\begin{figure}[] 
\centering
\includegraphics[width=\textwidth]{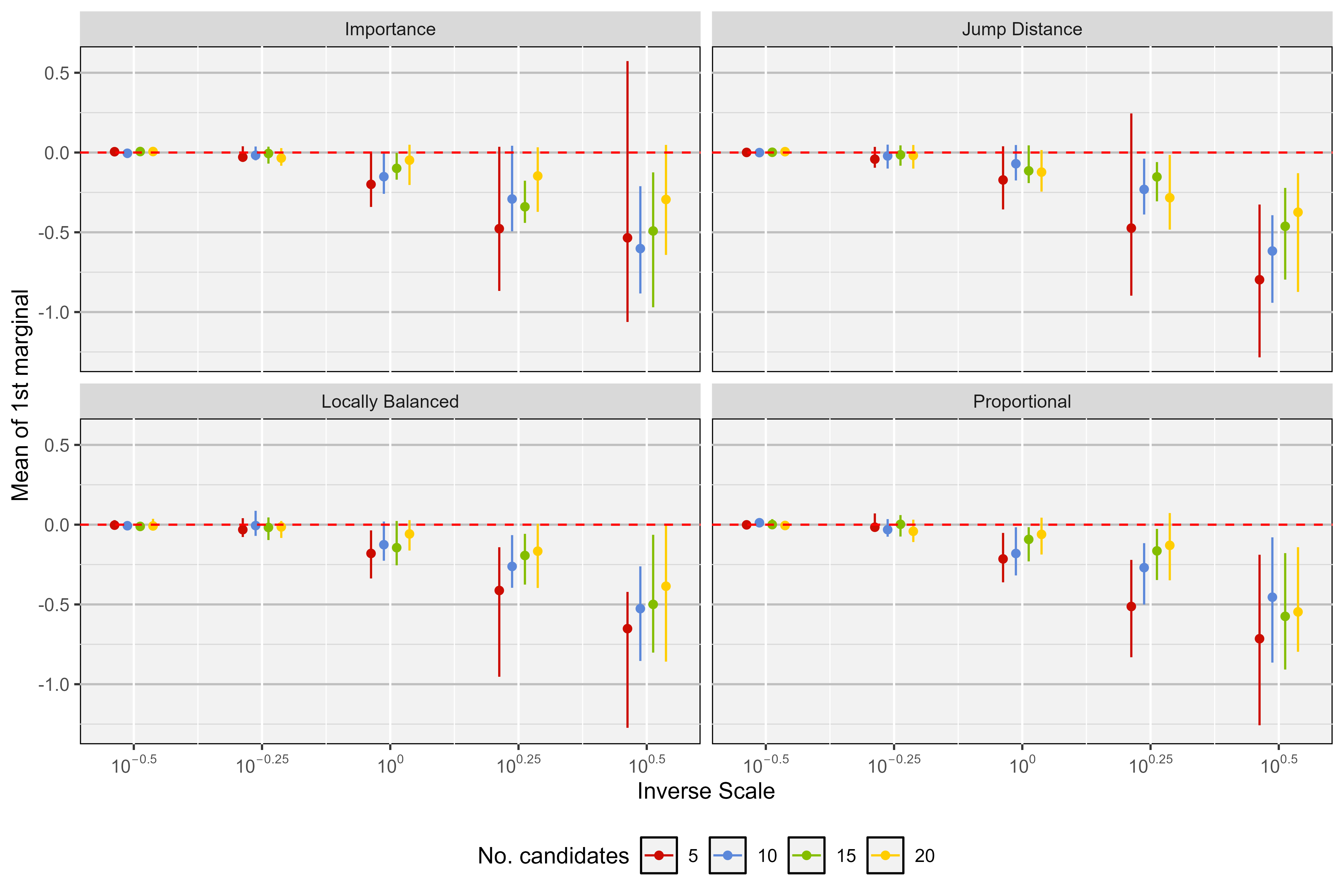}
\caption{Monte Carlo means of the first marginal ($y$) of Neal's funnel, generated by the Heterogeneous-CW proposal configuration. The true mean is shown by the red dashed line. Panels correspond to the weight function.}
\label{fig:funnel mean HeCW}
\end{figure}

The empirical means of $y$ are presented in Figure \ref{fig:funnel_mean}. Because the Monte Carlo error generated by the Heterogeneous-CW configuration is on a much smaller magnitude than the others, its results are shown separately in Figure \ref{fig:funnel mean HeCW}. As the funnel becomes steeper ($\beta^{-1}>1$), the random-walk kernels struggle to explore the narrow neck of the distribution, leading to biased mean estimates. Increasing $M$ tends to mitigates this bias. However, performance differs between proposal configurations: while Heterogeneous-CW maintains a consistently small Monte Carlo error across all $M$, the remaining three configurations exhibit an inflation in variance for $M \ge 15$.

\subsection{Multimodal Target Distribution}

Navigating multimodal landscapes and adequately sampling from all of their isolated high-density regions is challenging for many random-walk MCMC methods. To evaluate which MTM settings are best-suited for this task, our second experiment uses a classic benchmark: a $d$-dimensional, five-component Gaussian mixture model \citep{fontaine2022adaptive,modi2024DRHMC,yang2019componentwise}.
The mixture components are defined with unit covariance matrices and the following mean vectors: $\mathbf 0_d$, $\mathbf 3_d$, $-\mathbf 3_d$, $[3,-3,3,\ldots]^T_d$, and $[-3,3,-3,\ldots]^T_d$. The mixture weights for these five components are unbalanced, set at $[1,2,4,2,1]$. 

We want to assess how well each alorithm setting is able to recover these five modes as well as the appropriate mixture weights. Because many standard performance measures are not well-suited for evaluating multimodal sampling, we compare the MTM-generated chains to a baseline sample drawn directly from the target distribution. We compute the two-sample Kolmogorov-Smirnov distance (KSD) between the algorithm-generated samples and this baseline. We test each algorithm setting for dimension, $d \in \{2, 4, 6, 8, 10\}$, and $M \in \{1, 5, 10, 15, 20\}$ candidate draws. Crossed with the aforementioned weight functions and proposal configurations, this yields 400 experiment settings, each allocated a fixed wall-clock budget of 30 seconds. Because the automated mean-stabilization burn-in procedure (Algorithm S3) is ill-suited for multimodal posteriors, we instead compute the KSD under burn-in thresholds of 25\%, 50\%, and 75\%, retaining the minimum distance to conservatively estimate stationary sampling performance.

\begin{figure}[] 
\centering
\includegraphics[width=\textwidth]{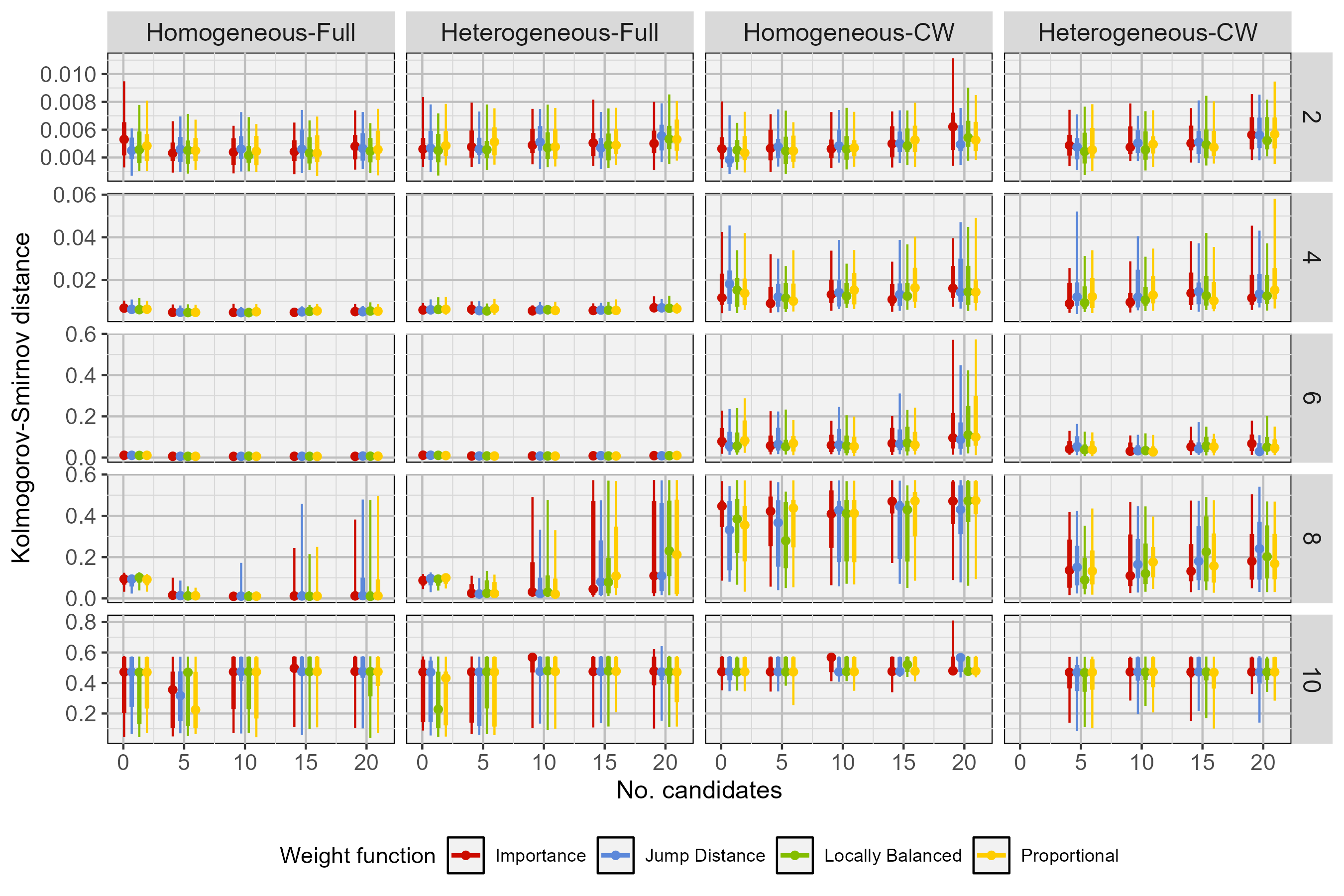}
\caption{Kolmogorov-Smirnov distance between MTM-generated samples and a baseline sample from the target. Results from 50 independent runs are shown by the empirical medians (points), 50\% intervals (thick lines), and 90\% intervals (thin lines). Column panels correspond to proposal configurations and row panels to the target dimension.}
\label{fig:KS_raw}
\end{figure}

The distribution of KSD across 50 independent runs for each setting are presented in Figure \ref{fig:KS_raw}. In the lowest-dimensional setting ($d=2$), all configurations exhibit comparable, strong performance. However, as the dimensionality increases to $d=4$ and $d=6$, the component-wise configurations begin to yield a slightly higher median KSD than the full-block proposals. This is accompanied by a substantial increase in across-run variance. At $d=8$, performance degrades sharply for the Homogeneous-CW configuration and moderately for the Heterogeneous-CW configuration. 
Conversely, the full-block proposals demonstrate a marginal but consistent improvement in median performance as $M$ increases beyond the MH baseline ($M=1$). Nevertheless, for larger candidate pools ($M=10$ or $15$), the variability in full-block performance increases, overlapping substantially with the component-wise settings. At $d=10$, the target becomes prohibitively difficult: all settings perform poorly, though the full-block schemes occasionally produce runs with adequate mode recovery. Consistent with our findings on non-Gaussian targets, the choice of weight function shows no impact on sample quality.

\subsection{Posterior Parameter Inference}

The final set of experiments evaluates the algorithms' performance on three Bayesian posterior distributions, where the goal is to estimate unknown model parameters. We use the Monte Carlo standard error (MCSE) of the posterior mean estimates of these parameters as the metric for comparing algorithm settings.
The first target is a synthetic Bayesian regression model, while the latter two are posteriors derived from real datasets. 

For the two real-data posteriors, a small fraction of the simulated Markov chains either failed to converge within the allotted time or became trapped in low-probability regions. To ensure the MCSE calculations reflect stationary sampling variation, we apply a second post-processing step, after the removal of burn-in, analogous to standard outlier detection. Any chain yielding a posterior mean outside 1.5 times the interquartile range of the distribution of means across all runs was classified as an outlier and discarded. Across all experimental settings, the maximum number of runs discarded for any single configuration was 6 out of 50 (12\%).

\paragraph*{Bayesian regression} The first target is a posterior distribution derived from a simulated multiple linear regression model:
\begin{equation}\label{eqn:regression}
y_i=\beta_0+\pmb\beta^T\mathbf x_i +\epsilon_i, \quad \epsilon_i\sim\mathcal N\left(0,\sigma^2\right),
\end{equation}
where $\mathbf x_i$ is a $d$-dimensional vector of predictor variables, $\beta_0$ is the intercept, $\pmb\beta$ is a $d$-dimensional vector of regression coefficients, and $\sigma$ is the observation noise. A dataset of $1000$ observations was generated using $d=4$, with true parameters $\beta_0=1$, $\pmb\beta = (0.1, 5, -5, 10)^T$, and $\sigma=0.5$. The Bayesian model imposes a Gaussian prior $\mathcal N(0, 100^2)$ for the intercept and each regression coefficient, and an inverse-gamma prior with mean 1 and variance 100 for the $\sigma^2$. 
We evaluate 96 experiment settings, crossing the previously defined proposal configurations and weight functions with candidate pool sizes $M \in \{1,2,4,6,8,10\}$. To account for variation in the observation process, we generate 10 independent datasets. For each dataset and algorithm configuration, 50 chains are run for 90 seconds. The standard deviation of the posterior means across these 50 chains provides a single MCSE estimate; the median of these 10 dataset-level MCSEs is reported.

\begin{figure}[] 
\centering
\includegraphics[width=\textwidth]{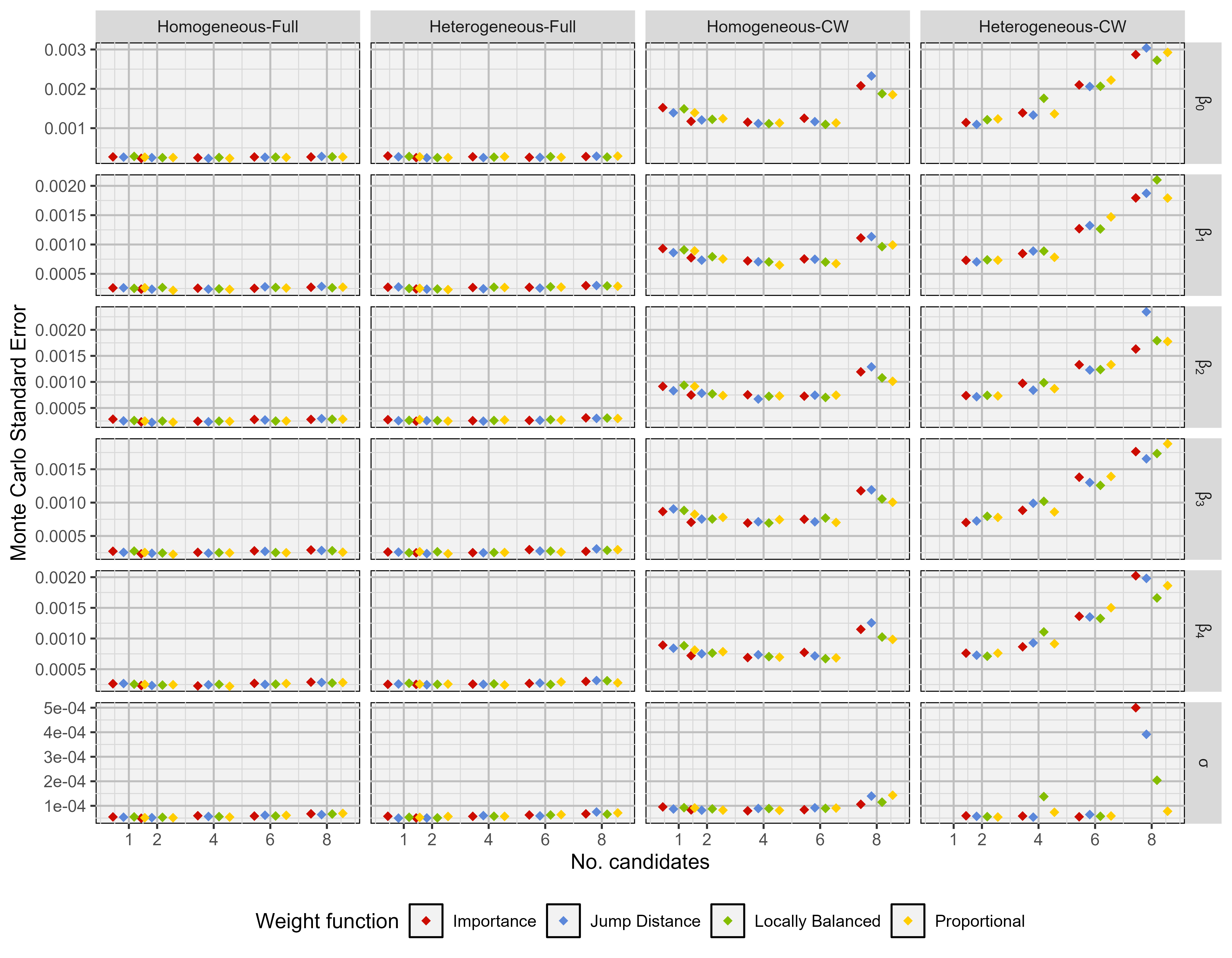}
\caption{Median Monte Carlo standard error (MCSE) of the posterior mean of the Bayesian regression parameters. Points represent the median over 10 randomly generated datasets, where each dataset's MCSE is calculated from 50 independent runs. Column panels correspond to proposal configurations and the row panels denote the parameter being estimated.}
\label{fig:MCSE all}
\end{figure}

The median MCSE for each parameter is visualized in Figure \ref{fig:MCSE all}. The full-block proposals produce substantially lower Monte Carlo error than the component-wise methods across all parameters. For the component-wise methods, the lowest Monte Carlo error is achieved at  $M=2$, with the error increasing or plateauing as the number of candidates increases. As the overall MCSE increases, the variation in performance attributable to the choice of weight function also increases, mostly visible in the component-wise results. For simulation results that have been disaggregated by dataset, see Figures S1--S4 in the Supplementary Material. 

\begin{figure}[] 
\centering
\includegraphics[width=\textwidth]{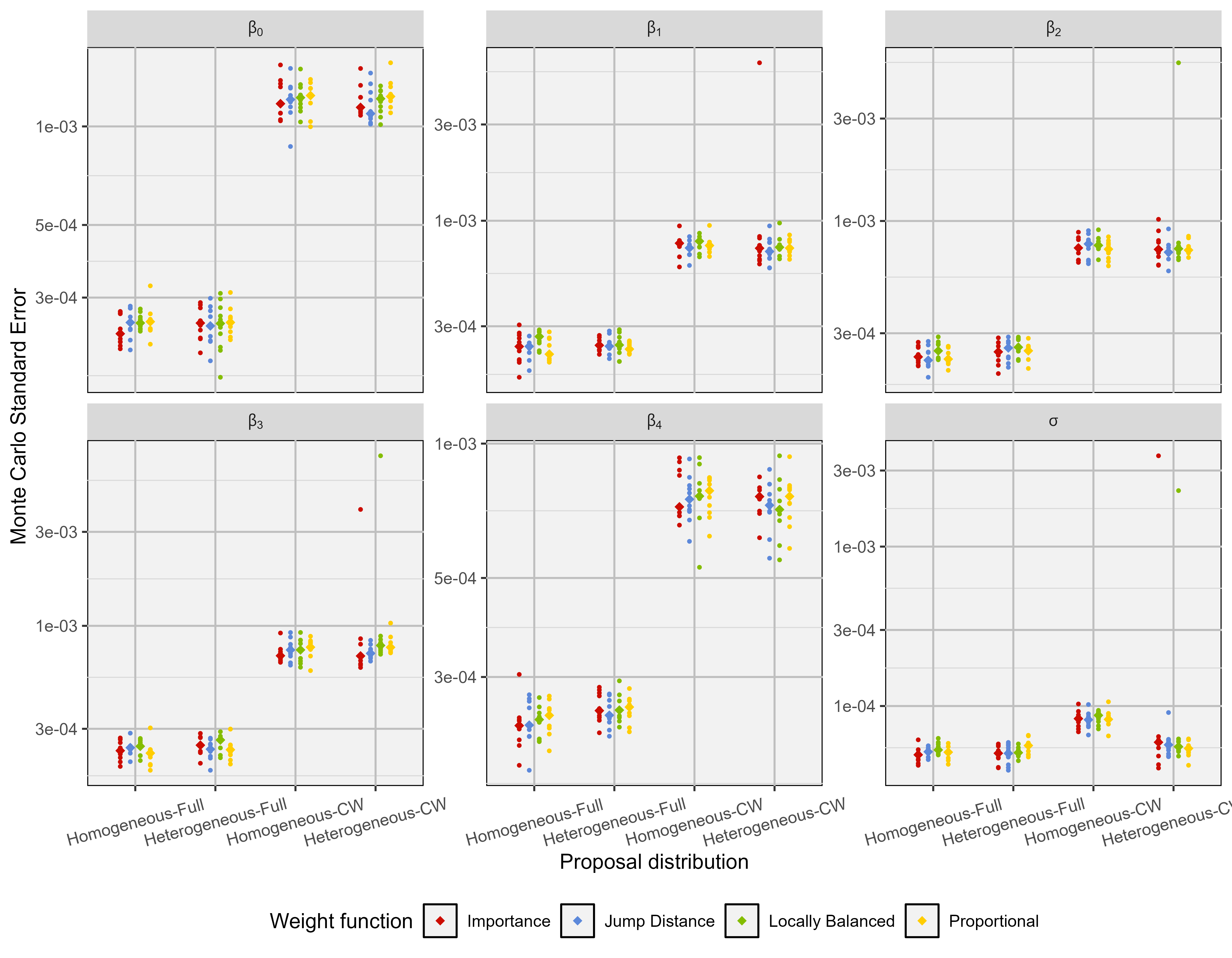}
\caption{Monte Carlo standard error (MCSE) for the Bayesian regression parameters fixed at $M=2$ candidates. Points represent the dataset-specific MCSE derived from 50 independent runs across 10 random datasets.}
\label{fig:MCSE M2}
\end{figure}

Because $M=2$ shows the lowest MCSE for all proposal configurations, Figure \ref{fig:MCSE M2} isolates the performance under this setting. The two full-block proposals perform similarly to one another, as do the two component-wise proposals, though Heterogeneous-CW exhibits greater cross-dataset variability than Homogeneous-CW. This pattern of performance is consistent between the regression coefficients. However, for the variance parameter, $\sigma$, the Homogeneous-CW setting exhibits the greatest MCSE while the other three settings are comparable. As in our previous experiments, no single weight function demonstrates a clear, systemic advantage.

\paragraph*{Gull's Lighthouse} Gull's lighthouse problem \citep{gull1988lighthouse} poses a posterior derived from a very small sample and heavy-tailed likelihoods \citep{modi2024DRHMC}. In this experiment, a boat observes flashes from a lighthouse that is located at position $x_0$ along the coastline at a distance $y$ from the boat; both $x_0$ and $y$ are unknown to the observer. Only three flashes are observed at locations $x_1, x_2$, and $x_3$ along the shore.  
The likelihood of observing $x_i$ given $x_0$ and $y$ is $L(x_i|x_0,y)=\frac{y}{\pi(y^2+(x_i-x_0)^2)}$, which is the probability density function of a $\text{Cauchy}(x_0,y)$ distribution at $x_i$. Imposing improper uniform priors on both $x_0$ and $y$, the target posterior distribution is: 
$$\pi(x_0,y|x_1,x_2,x_3)\propto\prod_{i=1}^3\text{Cauchy}(x_i|x_0,y).$$
We evaluated 96 settings using $M \in \{1, 10, 20, \dots, 50\}$ candidate draws. Each configuration was run for 20 seconds.

\begin{figure}[] 
\centering
\includegraphics[width=\textwidth]{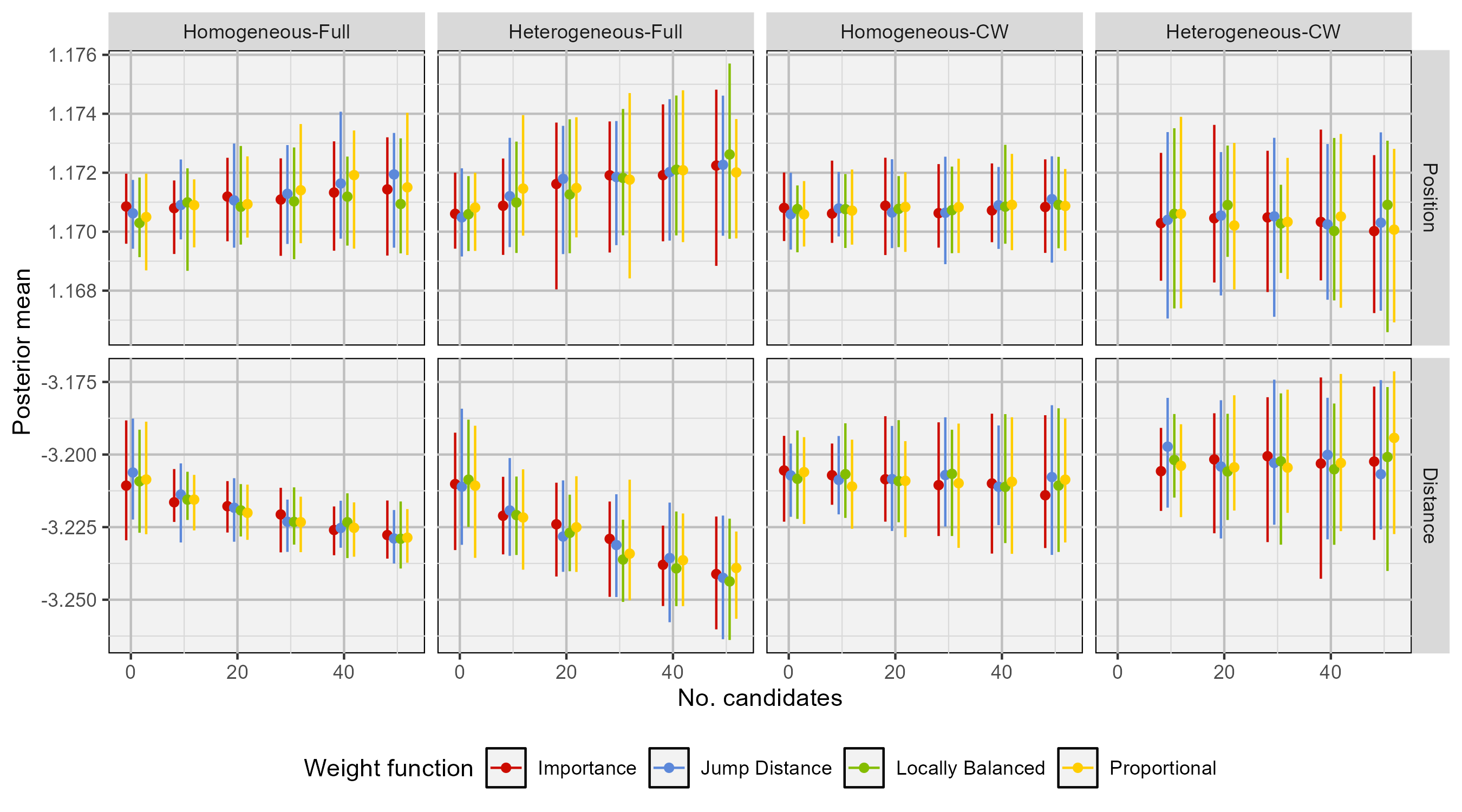}
\caption{Posterior mean estimates for position ($x_0$) and distance ($y$) in Gull's lighthouse problem. Points represent the median and vertical lines the 95\% interval across 50 independent runs. Column panels correspond to the proposal distribution configuration.}
\label{fig:lighthouse}
\end{figure}

The distribution of the posterior mean estimates of the lighthouse position ($x_0$) and distance ($y$) is shown in Figure \ref{fig:lighthouse}. The component-wise proposals produce estimates that are stable with respect to both the weight function and the number of candidates, though with a slight increase in the variability of $\hat y$ as $M$ increases. Full-block proposals exhibited different behaviour. While the estimates produced by MH ($M=1$) agree with the stable component-wise estimates, increasing the number of candidates induces a systematic shift in the Monte Carlo mean, producing an overestimate of the position and an underestimate of the distance. While there are differences in the point estimates of $y$ between Homogeneous-Full and Heterogeneous-Full, they are minor and do not demonstrate a statistically significant divergence between the two.

\paragraph*{Eight Schools} The eight schools problem is a canonical example of a Bayesian hierarchical model \citep{gelman2011BDA,rubin1981eightschools}, arising from a meta-analysis of SAT coaching effectiveness across eight high schools. For each school $i=1,\dots,8$, the observed treatment effect $y_i$ is modeled as a Gaussian random variable centered on a school-specific latent mean $\theta_i$, with standard error $\sigma_i$. These school-specific means share a hyperprior distribution with mean $\mu$ and standard deviation $\tau$. The posterior is defined by the following hierarchical model \citep{gelman2011BDA}:
\begin{align*}
    \mu &\sim \mathcal N(0,5^2), & \tau &\sim \text{Cauchy}_+(0,5), \\
    \theta_i&\sim\mathcal N(\mu,\tau^2), & y_i&\sim\mathcal N(\theta_i,\sigma_i^2).
\end{align*}
We tested 80 experimental settings ($M \in \{1,5,10,15,20\}$), each for a fixed runtime of 45 seconds.

\begin{figure}[] 
\centering
\includegraphics[width=\textwidth]{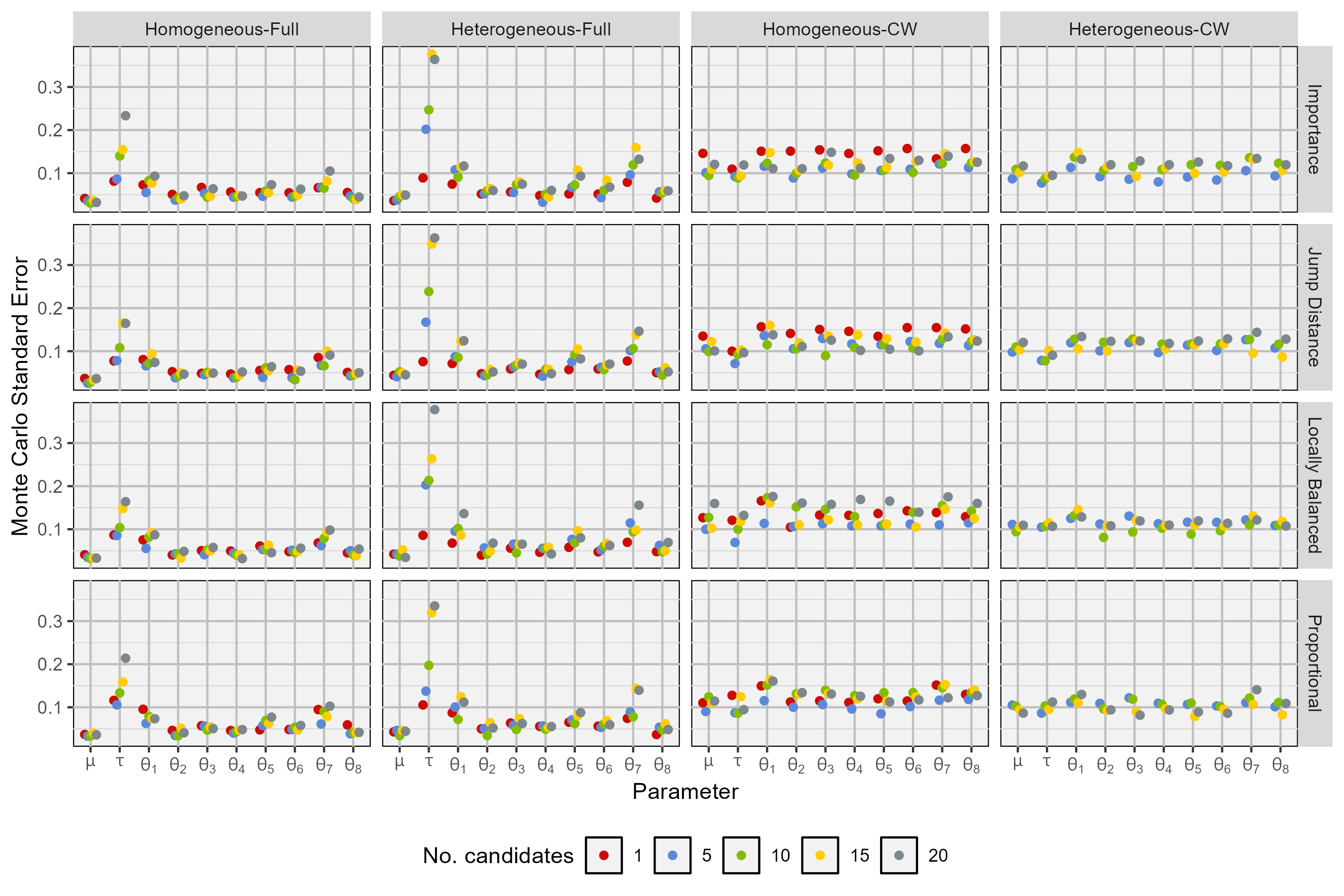}
\caption{Median Monte Carlo standard error of the posterior mean for the parameters in the eight schools hierarchical model. Column panels correspond to the proposal configurations and row panels to the weight function.}
\label{fig:eight schools}
\end{figure}

The median MCSEs of the posterior means of all parameters are presented in Figure \ref{fig:eight schools}. The component-wise proposals exhibit similar variability for all parameters with the choice of weight function having negligible impact. The full-block proposals tend to produce estimates with lower variability than the component-wise proposals with the exception of the global scale parameter $\tau$ and for specific school means ($\theta_1$ and $\theta_7$). In most settings, increasing the number of candidates yields a slight increase in MCSE, especially in the estimation of $\tau$.

\section{Discussion}
\label{sec:discussion P1}

Multiple-try Metropolis is a local MCMC algorithm that uses a multi-candidate proposal mechanism to improve the efficiency of its transition kernel \citep{liu2000mtm}. Since its inception, several authors have contributed to its methodological development. In this paper, we collated many of these developments, present them within a unified MTM framework, and explore their trade-offs through simulation.

Motivated by involutive MCMC \citep{neklyudov2020involutivemcmc}, we introduced a constructive procedure for deriving valid acceptance probabilities for reversible Metropolis algorithms. This procedure takes any proposal mechanism and returns an acceptance ratio; a transition kernel comprised of these two components is guaranteed to satisfy the detailed balance condition. We present this as a general tool for designing a wide variety of novel MCMC algorithms. MTM was treated as a use-case, recovering the general form for the MTM acceptance probability.

The MTM literature emphasizes the development of novel weight functions, which is often coincidental with development of the proposal distribution(s). Consequently, separating the effect of the novel weight function from that of the novel proposal distribution is seldom possible in previous experiments. By using a factorial design for our simulation study, we were able to prise apart these effects and found that algorithm performance is driven primarily by the proposal configuration and the number of candidate draws, rather than the weight function.
The negligible impact of the weight function observed in our experiments is in contrast with its prevalence in the MTM literature.
In particular, the locally balanced weight function is supported by theoretical results \citep{gagnon2023localbalancing}. However, these results primarily concern convergence rates and pre-stationary behavior. Our experiment evaluates the stationary performance of the algorithms and so does not reflect this. Therefore, while our findings demonstrate that the choice of weight function has little effect on sampling efficiency once stationarity is reached, this does not preclude its potential importance in accelerating convergence or reducing burn-in for some targets.

Regarding the effect of the number of candidates, an experiment from a previous study of MTM found that increasing the number of candidates improved performance \citep{martino2018review}. By comparing raw ESS to ESS per iteration in our experiments (see Figures \ref{fig:banana_raw} and \ref{fig:banana_std}), we find that the gains in sampling efficiency from increasing the number of candidates is outweighed by the resulting increase in computation time. The full benefit of increasing the number of candidates is best achieved by using parallel computing regimes to reduce or eliminate this additional computational burden.

It was the proposal configuration that was the greatest determinant of performance in the simulation study. We provide the following guidelines for selecting a proposal configuration based on features of the target distribution, informed by the results of our study. For highly non-Gaussian targets, the component-wise proposal paired with a balanced selection adaptation exhibited the best performance, maximizing both per-second and per-iteration effective sample sizes. Conversely, in the presence of multimodality, the component-wise configurations struggled with navigating between modes compared to their full-block counterparts. 
For the relatively simple parameter estimation problem presented by the Bayesian regression posterior, full-block proposals achieved lower and more stable MCSEs. However, posteriors derived from real data presented more challenging targets, exhibiting heavy tails in Gull's lighthouse problem and a funnel-shaped posterior in the eight schools hierarchical model. The component-wise proposals yielded stable point estimates in both cases, whereas full-block proposals were more sensitive to the number of candidates. These benchmarks suggest the following approach for designing the proposal distribution: if your target distribution exhibits non-Gaussian features, then a component-wise proposal is likely to be better than a full-block proposal; if your target is reasonably Gaussian or exhibits relatively low-dimensional multimodality, then a full-block proposal more likely to perform better.


\paragraph{Acknowledgements} This work was funded by the Natural Sciences and Engineering Research Council of Canada, Discovery grant (RGPIN-2019–06131). Several discussions with Alexandre Bouchard-C\^ot\'e were very helpful in developing the procedure outlined in Section \ref{sec:derivation P1} and in improving the clarity of the manuscript.

\bibliographystyle{unsrtnat}
\bibliography{references}

\section*{Supplementary Materials}

\makeatletter
\renewcommand{\thefigure}{S\@arabic\c@figure}
\makeatother
\setcounter{figure}{0}

\makeatletter
\renewcommand{\thealgorithm}{S\@arabic\c@algorithm}
\makeatother
\setcounter{algorithm}{0}

\makeatletter
\renewcommand{\thesection}{S\@arabic\c@section}
\makeatother
\setcounter{section}{0}

This supplementary materials file is organized into the following sections:
\begin{itemize}
    \item [Section S1:] Additional details and pseudocode for the adaptation algorithms referred to in Section 3.1 of the manuscript.
    \item [Section S2:] Pseudocode for the automatic burn-in detection algorithm described in Section 4 of the manuscript.
    \item [Section S3:] Additional results from the simulation study conducted in Section 4 of the manuscript.
\end{itemize}

\section{Adaptation algorithms for proposal distributions}

\subsection{``Adaptive Metropolis'' for Homogeneous Proposals}

What we term the ``adaptive Metropolis'' algorithm in the manuscript is a well-established procedure for automatic tuning of the proposal covariance, $\Sigma$ \citep{andrieu2008,fontaine2022adaptive}.
For a chain at state $\mathbf x$, the proposals are generated from a Gaussian random walk, $\mathcal N(\mathbf x,\Sigma\times2.38^2/d)$. This factor of $2.38^2/d$ is an optimal scaling parameter for these types of adaptation algorithms \citep{roberts1997optimalscaling}. An initial value of $\Sigma_0$ is used for the first $n_0$ iterations. After this initial stage, the covariance is updated using the current state of the chain. Given that $\pmb\mu_n=\pmb\mu_0$ (some initial value) and $\Sigma_n=\Sigma_0$ for $n\le n_0$, the proposal covariance is updated by the following formulae at iteration $n$:
\begin{align}\label{eqn:am update}
\pmb\mu_n &= \pmb\mu_{n-1} + \gamma(n)[x_n-\pmb\mu_{n-1}] \\
\Sigma_n &= \Sigma_{n-1}+\gamma(n)\left[(x_n-\pmb\mu_n)(x_n-\pmb\mu_n)^T-\Sigma_{n-1}\right].
\end{align}
The scaling term, $\gamma(n)$, is a learning rate that decreases with $n$. This decreasing learning rate is used to preserve the ergodicity of the chain. In our implementation we use $\gamma(n)=n^{-0.6}$ \citep{fontaine2022adaptive}. The pseudocode for this algorithm is given in Algorithm \ref{algo:Adaptive metropolis}.

\begin{algorithm}[h!]
\caption{Adaptive Metropolis Covariance Update}
\label{algo:Adaptive metropolis}
\begin{algorithmic}[1]
\State \textbf{Input:} Current state $\mathbf x_n$; current iteration $n$; current proposal covariance $\Sigma$ and running mean $\pmb\mu$.
\State \textbf{Settings:} Burn-in threshold $n_0 = 100$.
\If{$n \le n_0$}
    \State \Return $\Sigma, \pmb\mu$ \Comment{No adaptation during initial burn-in}
\Else
    \State Compute the decaying learning rate $\gamma = n^{-0.6}$.
    \State Update the running mean: $\pmb\mu = \pmb\mu + \gamma(\mathbf x_n - \pmb\mu)$.
    \State Update the proposal covariance: $\Sigma = \Sigma + \gamma \left( (\mathbf x_n - \pmb\mu)(\mathbf x_n - \pmb\mu)^T - \Sigma \right)$.
    \State \Return $\Sigma, \pmb\mu$
\EndIf
\end{algorithmic}
\end{algorithm}

\subsection{``Balanced Selection Rate'' Adaptation for Heterogeneous Component-Wise MTM}

This adaptation algorithm has been design specifically for MTM \citep{yang2019componentwise}. A sequence of increasing proposal variances, $\sigma_{i,m}^2$, is used to generate proposals corresponding to both large and small jumps in the state space. To ensure efficient exploration, one desires that proposals from both large and small jumps are accepted. Therefore, this adaptation scheme was designed to modify the extreme values of this variance sequence to preserve a balance in the selection rates. We define the selection rate of the $m$-th candidate for coordinate $i$ as the number of times that candidate $m$ was selected as the proposal, divided by the number of iterations since the last time an update was performed. Adaptation is only triggered every $\beta$ iterations and when triggered, the actual adaption is only carried out with probability $P(n)=\max(0.99^{a-1},a^{-1/2})$, where $a=(n-\beta)/\beta$. If candidates generated from the largest variance ($\sigma_{i,M}$) are selected too frequently, the upper bound is doubled; if they are rarely selected, it is halved. Symmetric conditions apply to the smallest variance ($\sigma_{i,1}$). Whenever the bounds are updated, the intermediate variances $\sigma_{i,2:M-1}$ are reset so that all $M$ standard deviations remain equally spaced on the $\log_2$ scale.

\begin{algorithm}[h!]
\caption{Balanced Selection Rate Adaptation}
\label{algo:balanced selection}
\begin{algorithmic}[1]
\State \textbf{Input:} Current iteration $n$; current scalar proposal variances $\{\sigma_{i,m}\}$; recorded candidate selection rates $S_{i,m}$.
\State \textbf{Settings:} Adaptation interval $\beta=100$; bounds $\epsilon=-15, L=50$.
\State \textbf{Require:} Number of candidates $M$, State dimension $d$.
\State Compute adaptation probability: $a = (n-\beta)/\beta$, $P = \max(0.99^{a-1}, a^{-1/2})$.
\State Draw $U \sim \text{Unif}(0,1)$.
\If{$(n \mod \beta = 0)$ \textbf{and} $(U < P)$}
    \For{$i = 1, \dots, d$}
        \State \textit{// Adjust upper bound}
        \If{$S_{i,M} > 2/M$}
            \State $\sigma_{i,M} = \min(2\sigma_{i,M}, 2^L)$.
            \State Rescale $\sigma_{i,1:M}$ to be equally spaced on the $\log_2$-scale.
        \ElsIf{$(S_{i,M} < 1/(2M))$ \textbf{and} $(\sigma_{i,M}/2 > \sigma_{i,1})$}
            \State $\sigma_{i,M} = \max(\sigma_{i,M}/2, 2^\epsilon)$.
            \State Rescale $\sigma_{i,1:M}$ to be equally spaced on the $\log_2$-scale.
        \EndIf
        
        \State \textit{// Adjust lower bound}
        \If{$S_{i,1} > 2/M$}
            \State $\sigma_{i,1} = \max(\sigma_{i,1}/2, 2^\epsilon)$.
            \State Rescale $\sigma_{i,1:M}$ to be equally spaced on the $\log_2$-scale.
        \ElsIf{$(S_{i,1} < 1/(2M))$ \textbf{and} $(2\sigma_{i,1} < \sigma_{i,M})$}
            \State $\sigma_{i,1} = \min(2\sigma_{i,1}, 2^L)$.
            \State Rescale $\sigma_{i,1:M}$ to be equally spaced on the $\log_2$-scale.
        \EndIf
    \EndFor
\EndIf
\State \Return the updated $\{\sigma_{i,m}\}$
\end{algorithmic}
\end{algorithm}

\section{Pseudocode for the automatic burn-in detection}
\begin{algorithm}[H]
\caption{\bf{Automatic burn-in}}
\label{algo:auto burnin}
{\fontsize{10pt}{10pt}\selectfont
\begin{algorithmic}[1]
\State {\bfseries Input:} Monte Carlo sample $\{\mathbf x_n\}_{n=1}^N$. 
\State Split chain into 20 equally-sized blocks: $n_j=j\times \lfloor N/20\rfloor$, $j=0,\ldots,20$.
\State Initialize sample to final block: $\mathcal X=\{\mathbf x_n\}_{n_{19}}^{N}$, $N_0=n_{19}$.
\State Compute $(L(\mathcal X),U(\mathcal X))$ as the empiric 95\% interval of $\mathcal X$.
\For{$j \in \{18,\ldots,1\}$}
    \State Compute mean of next block $\mu=\frac{1}{\lfloor N/20\rfloor}\sum_{i=n_{j-1}}^{n_j}\mathbf x_i$.
    \If{$\mu\in(L(\mathcal X),U(\mathcal X))$}
        \State Update our sample: $\mathcal X = \{\mathbf x_n\}_{n_{j-1}}^{N}$.
        \State Update the 95\% interval: $(L(\mathcal X),U(\mathcal X))$.
        \State Update the burn-in index: $N_0 = n_{j-1}$.
    \EndIf
    \If{$\mu\notin(L(\mathcal X),U(\mathcal X))$ for 2 consecutive samples}
        \State Return burn-in $N_0$ and sample $\mathcal X$.
    \EndIf
\EndFor

\end{algorithmic}
}
\end{algorithm}

\section{Additional results for the simulation study}

This section of the Supplementary Materials shows the Monte Carlo standard errors for each of the parameters in the Bayesian regression problem (Section 4.3). To better illustrate the differences between each of the proposal configurations, each figure shows the results for a different proposal configuration.


\begin{figure}[H] 
\centering
\includegraphics[width=\textwidth]{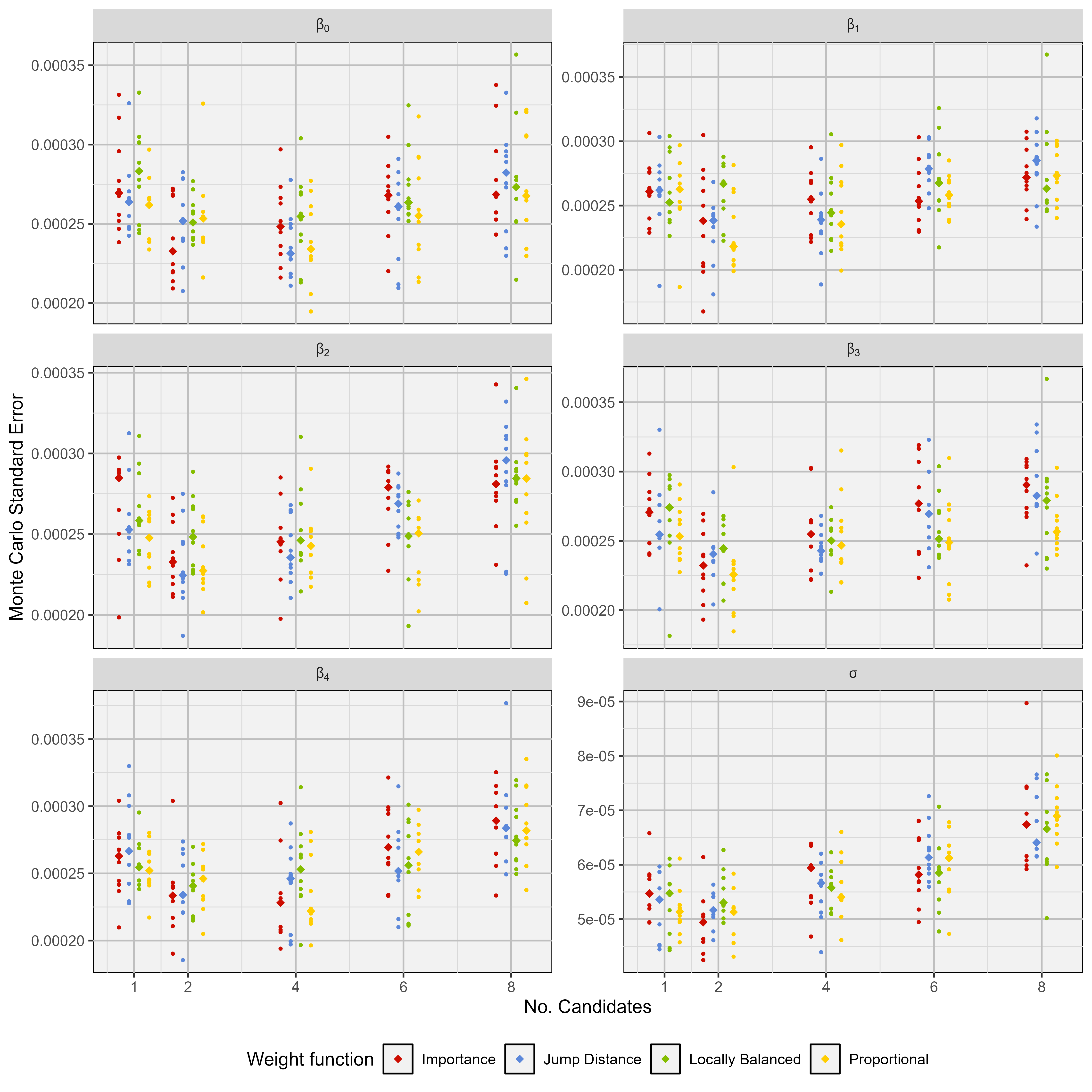}
\caption{Monte Carlo standard errors for the Homogeneous-Full proposal distribution. Small points are the MCSE for each data seed and large points are the median MCSE over all seeds. Results for $M=10$ are omitted.}
\label{fig:MCSE SF}
\end{figure}

\begin{figure}[H] 
\centering
\includegraphics[width=\textwidth]{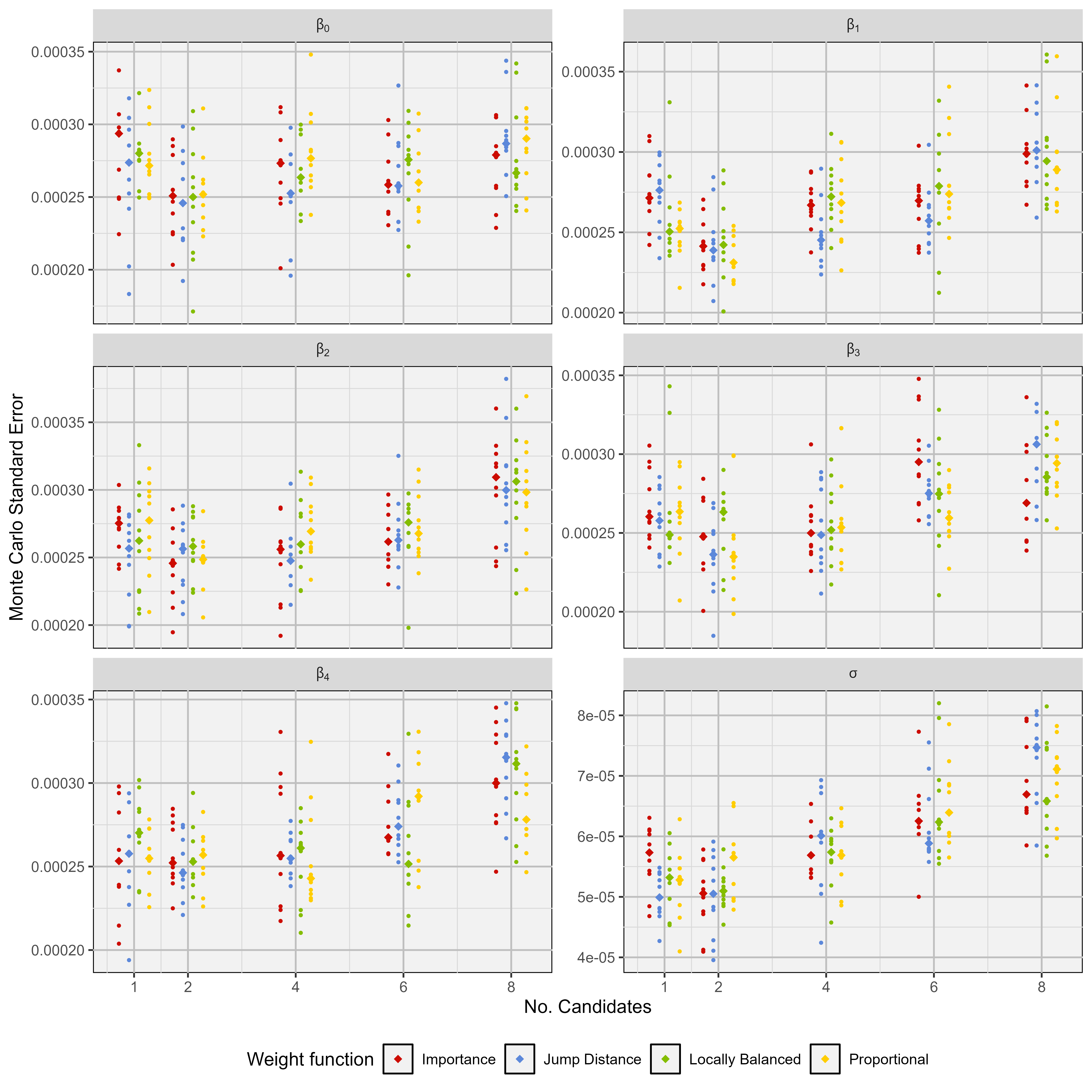}
\caption{Monte Carlo standard errors for the Heterogeneous-Full proposal distribution. Small points are the MCSE for each data seed and large points are the median MCSE over all seeds. Results for $M=10$ are omitted.}
\label{fig:MCSE MF}
\end{figure}

\begin{figure}[H] 
\centering
\includegraphics[width=\textwidth]{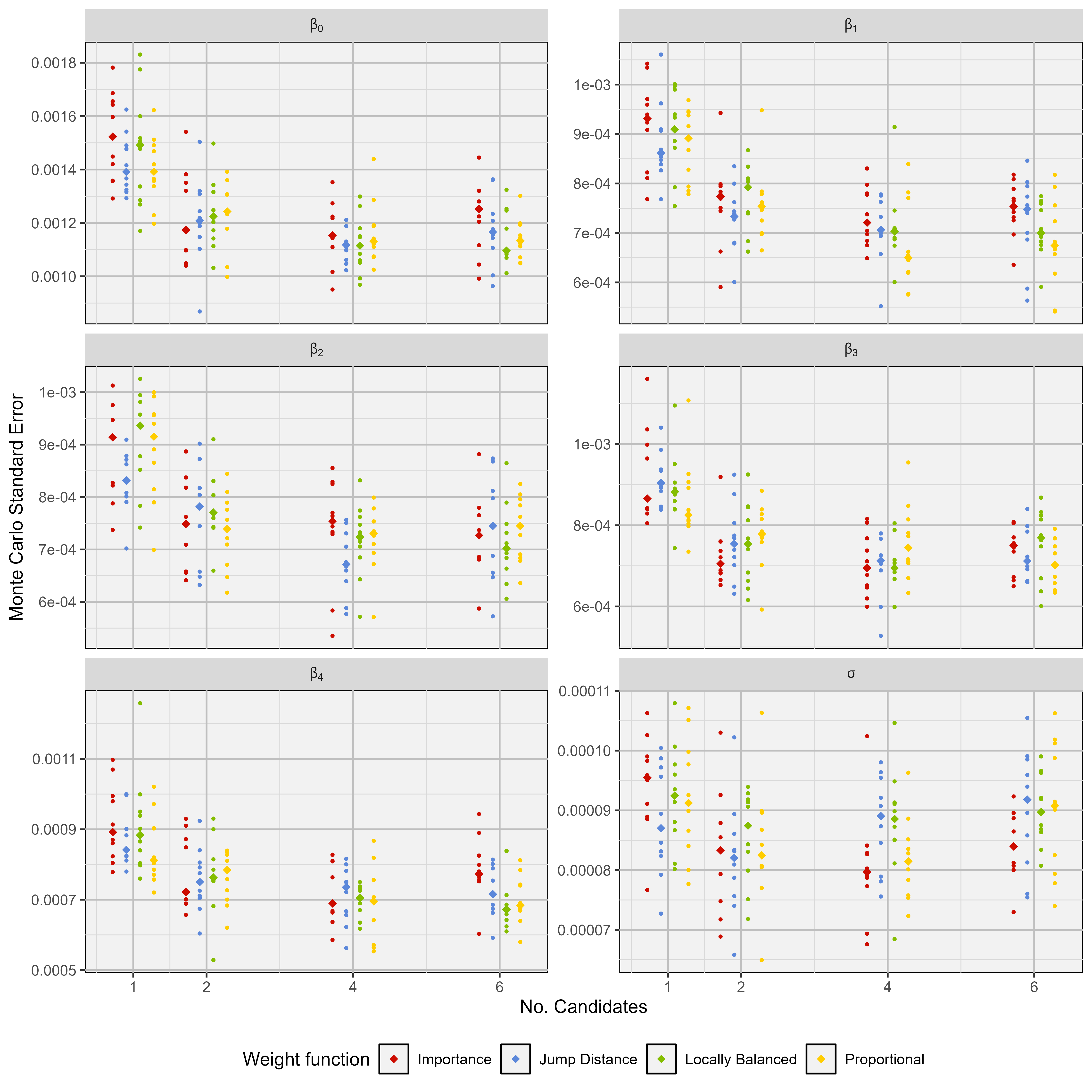}
\caption{Monte Carlo standard errors for the Homogeneous-CW proposal distribution. Small points are the MCSE for each data seed and large points are the median MCSE over all seeds. Results for $M=8,10$ are omitted.}
\label{fig:MCSE SCW}
\end{figure}

\begin{figure}[H] 
\centering
\includegraphics[width=\textwidth]{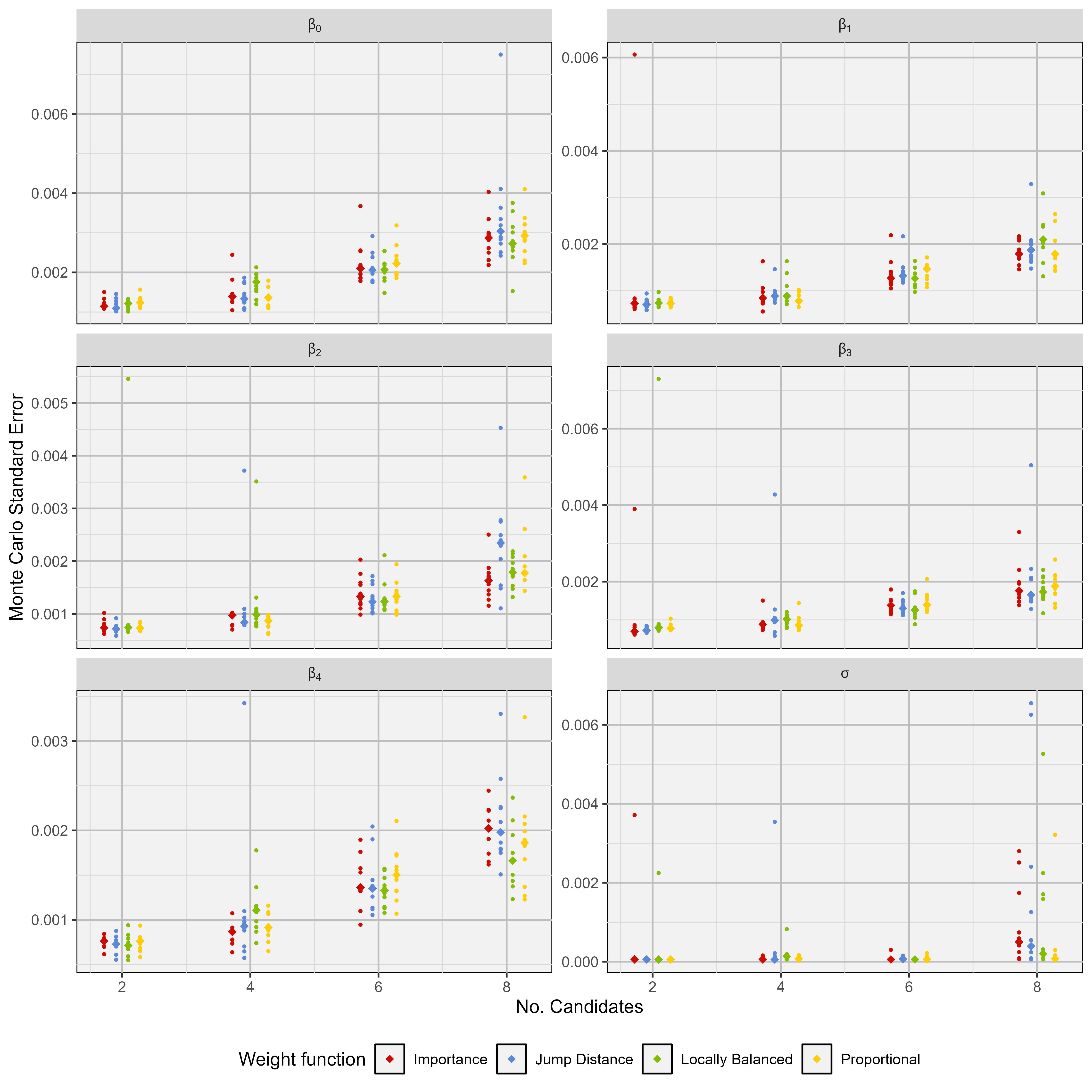}
\caption{Monte Carlo standard errors for the Heterogeneous-CW proposal distribution. Small points are the MCSE for each data seed and large points are the median MCSE over all seeds. Results for $M=10$ are omitted.}
\label{fig:MCSE MCW}
\end{figure}

\end{document}